\RequirePackage{ifpdf}
\ifpdf 
\documentclass[pdftex]{sigma}
\else
\documentclass{sigma}
\fi

\usepackage[all]{xy}     

\newcommand{\rrbr}{]\!]}
\newcommand{\llbr}{[\![}

\begin{document}

\allowdisplaybreaks

\renewcommand{\thefootnote}{$\star$}

\renewcommand{\PaperNumber}{064}

\FirstPageHeading

\ShortArticleName{Hochschild Homology and Cohomology of Klein Surfaces}

\ArticleName{Hochschild Homology and Cohomology\\ of Klein
Surfaces\footnote{This paper is a
contribution to the Special Issue on Deformation Quantization. The
full collection is available at
\href{http://www.emis.de/journals/SIGMA/Deformation_Quantization.html}{http://www.emis.de/journals/SIGMA/Deformation\_{}Quantization.html}}}

\Author{Fr\'ed\'eric BUTIN}

\AuthorNameForHeading{F. Butin}

\Address{Universit\'e de Lyon, Universit\'e Lyon 1, CNRS,
UMR5208, Institut Camille Jordan,\\ 43 blvd du 11 novembre 1918, F-69622 Villeurbanne-Cedex, France}
\Email{\href{mailto:butin@math.univ-lyon1.fr}{butin@math.univ-lyon1.fr}}
\URLaddress{\url{http://math.univ-lyon1.fr/~butin/}}

\ArticleDates{Received April 09, 2008, in f\/inal form September 04,
2008; Published online September 17, 2008}

\Abstract{Within the framework of deformation quantization, a
f\/irst step towards the study of star-products is the calculation
of Hochschild cohomology. The aim of this article is precisely to
determine the Hochschild homology and cohomology in two cases of
algebraic varieties. On the one hand, we consider singular curves
of the plane; here we recover, in a~dif\/ferent way, a result proved
by Fronsdal and make it more precise. On the other hand, we are
interested in Klein surfaces. The use of a complex suggested
 by Kontsevich and the help of Groebner bases allow us to solve the problem.}


\Classification{53D55; 13D03; 30F50; 13P10}

\renewcommand{\thefootnote}{\arabic{footnote}}
\setcounter{footnote}{0}

\section{Introduction}\label{sec1}

\subsection{Deformation quantization}\label{sec1.1}

\noindent Given a mechanical system $(M,\,\mathcal{F}(M))$, where
$M$ is a Poisson manifold and $\mathcal{F}(M)$ the algebra of
regular functions on $M$, it is important to be able to quantize
it, in order to obtain more precise results than through classical
mechanics.
An available method is deformation quantization, which
consists of constructing a star-product on the algebra of formal
power series $\mathcal{F}(M)[[\hbar]]$. The f\/irst approach for
this construction is the computation of
Hochschild cohomology of $\mathcal{F}(M)$.

 We consider such a mechanical system given by a~Poisson
manifold $M$, endowed with a~Poisson bracket $\{\cdot,\cdot\}$. In
classical mechanics, we study the (commutative) algebra
$\mathcal{F}(M)$ of regular functions (i.e., for example,
$\mathcal{C}^\infty$, holomorphic or polynomial) on $M$, that is
to say the observables of the classical system. But quantum
mechanics, where the physical system is described by a~(non
commutative) algebra of operators on a Hilbert space, gives more
correct results than its classical analogue. Hence the importance
to get a quantum description of the classical system
$(M,\,\mathcal{F}(M))$, such an operation is called a
quantization.

One option is geometric quantization, which allows us to construct
in an explicit way a~Hilbert space and an algebra of operators on
this space (see the book \cite{GRS07} on the Virasoro group and
algebra for a nice introduction to geometric quantization). This
very interesting method presents the drawback of being seldom
applicable.

That is why other methods, such as asymptotic
quantization and deformation quantization, have been introduced.
The latter, described in 1978 by F.~Bayen, M.~Flato, C.~Fronsdal,
A.~Lichnerowicz and D.~Sternheimer in~\cite{BFFLS78}, is a~good
alternative: instead of constructing an algebra of operators on a
Hilbert space, we def\/ine a formal deformation of $\mathcal{F}(M)$.
This is given by the algebra of formal power series
$\mathcal{F}(M)[[\hbar]]$, endowed with some associative, but not
always commutative, star-product,
\begin{gather}\label{def}
  f\ast
g=\sum_{j=0}^\infty m_j(f,\,g)\hbar^j,
\end{gather}
where the maps $m_j$ are bilinear and where $m_0(f,\,g)=fg$. Then
quantization is given by the map $f\mapsto \widehat{f}$, where the
operator $\widehat{f}$ satisf\/ies $\widehat{f}(g)=f\ast g$.

 In which cases does a Poisson manifold admit such a quantization? The answer was given by Kontsevich in \cite{K97}: in fact he
constructed a star-product on every Poisson manifold. Besides, he
proved that if $M$ is a smooth manifold, then the equivalence
classes of formal deformations of the zero Poisson bracket are in
bijection with equivalence classes of star-products. Moreover, as
a consequence of the Hochschild--Kostant--Rosenberg
theorem, every Abelian star-product is equivalent to a trivial one.

In the case where $M$ is a singular algebraic variety, say
\[
M=\{\mathbf{z}\in \mathbb{C}^n \, /\,  f(\mathbf{z})=0\},
\]
with $n=2$ or $3$, where $f$ belongs to $\mathbb{C}[\mathbf{z}]$
-- and this is the case which we shall study -- we shall consider
the algebra of functions on $M$, i.e.\ the quotient algebra
$\mathbb{C}[\mathbf{z}]\,/\,\langle f\rangle$. So the above
mentioned result is not always valid. However, the deformations of
the algebra $\mathcal{F}(M)$, def\/ined by the formula~(\ref{def}),
are always classif\/ied by the Hochschild cohomology of
$\mathcal{F}(M)$, and we are led to the study of the Hochschild
cohomology of $\mathbb{C}[\mathbf{z}]\,/\,\langle
f\rangle$.

\subsection{Cohomologies and quotients of polynomial algebras}\label{sec1.2}

\noindent We shall now consider $R:=\mathbb{C}[z_1, \dots,
z_n]=\mathbb{C}[\mathbf{z}]$, the algebra of polynomials in $n$
variables with complex coef\/f\/icients. We also f\/ix $m$ elements
$f_1, \dots, f_m$ of $R$, and we def\/ine the quotient algebra
$A:=R\, /\, \langle f_1, \dots, f_m\rangle=\mathbb{C}[z_1,
\dots, z_n]\, /\,
\langle f_1, \dots, f_m\rangle$.

Recent articles were devoted to the study of particular
cases, for Hochschild as well as for Poisson homology and cohomology:

\begin{enumerate}\itemsep=0pt
\item[] C. Roger and P. Vanhaecke, in \cite{RV02},
 calculate the Poisson cohomology of the af\/f\/ine plane~$\mathbb{C}^2$, endowed
with the Poisson bracket
$f_1\,\partial_{z_1}\wedge\partial_{z_2}$, where $f_1$ is a
homogeneous polynomial. They express it in terms of the number of
irreducible components of the singular locus $\{\mathbf{z}\in
\mathbb{C}^2\, /\, f_1(\mathbf{z})=0\}$ (in this case, we have a
symplectic structure outside the singular locus), the algebra of
regular functions on this curve being the quotient algebra
$\mathbb{C}[z_1,\,z_2]\,/\,\langle f\rangle$.

\item[]
M.~Van den Bergh and A.~Pichereau, in \cite{VB94,P05} and \cite{P06}, are interested in the case where $n=3$
and $m=1$, and where $f_1$ is a weighted homogeneous polynomial
with an isolated singularity at the origin. They compute the
Poisson homology and cohomology, which in particular may be
expressed in terms of the Milnor number of the space
$\mathbb{C}[z_1,\,z_2,\,z_3]\, /\, \langle
\partial_{z_1}f_1,\,\partial_{z_2}f_1,\,\partial_{z_3}f_1\rangle$ (the def\/inition
of this number is given in~\cite{AVGZ86}).

\item[]
Once more in the case where $n=3$ and $m=1$, in
\cite{AL98}, J.~Alev and T.~Lambre compare the Poisson homology in
degree~$0$ of Klein surfaces with the Hochschild homology in
degree~$0$ of $A_1(\mathbb{C})^G$, where $A_1(\mathbb{C})$ is the
Weyl algebra and $G$ the group associated to the Klein surface. We
shall give more details about those surfaces in Section~\ref{sec4.1}.

\item[] In \cite{AFLS00}, J.~Alev, M.A. Farinati,
Th.~Lambre and A.L.~Solotar establish a fundamental
result: they compute all the Hochschild homology and cohomology
spaces of $A_n(\mathbb{C})^G$, where $A_n(\mathbb{C})$ is the Weyl
algebra, for every f\/inite subgroup $G$ of
$\mathbf{Sp}_{2n}\mathbb{C}$. It is an interesting and classical
question to compare the Hochschild homology and cohomology of
$A_n(\mathbb{C})^G$ with the Poisson homology and cohomology of
the ring of invariants $\mathbb{C}[\mathbf{x},\,\mathbf{y}]^G$,
which is a quotient algebra of the
form~$\mathbb{C}[\mathbf{z}]\,/\,\langle f_1,\dots,f_m\rangle$.

\item[] C.~Fronsdal studies in \cite{FK07} Hochschild homology
and cohomology in two particular cases: the case where $n=1$ and
$m=1$, and the case where $n=2$ and $m=1$. Besides, the appendix
of this article gives another way to calculate the Hochschild
cohomology in the more general case of complete intersections.
\end{enumerate}

In this paper, we propose to calculate the Hochschild
homology and cohomology  in two particularly important cases.

\begin{itemize}
\item The case of singular curves of the plane, with
polynomials $f_1$ which are weighted homogeneous polynomials with
a singularity of modality zero: these polynomials correspond to
the normal forms of weighted homogeneous functions of two
variables and of modality zero, given in the classif\/ication of
weighted homogeneous functions of \cite{AVGZ86} (this case already
held C. Fronsdal's attention).

\item The case of Klein surfaces $\mathcal{X}_\Gamma$ which
are the quotients $\mathbb{C}^2\,/\,\Gamma$, where $\Gamma$ is a
f\/inite subgroup of $\mathbf{SL}_2\mathbb{C}$ (this case
corresponds to $n=3$ and~$m=1$). The latter have been the subject
of many works; their link with the f\/inite subgroups of
$\mathbf{SL}_2\mathbb{C}$, with the Platonic polyhedra, and with
McKay correspondence explains this large interest. Moreover, the
preprojective algebras, to which \cite{CBH98} is devoted,
constitute a family of deformations of the Klein surfaces,
parametrized by the group which is associated to them: this fact
justif\/ies once again the
calculation of their cohomology.
\end{itemize}

The main result of the article is given by two propositions:

\begin{proposition}
Given a singular curve of the plane, defined by a polynomial
$f\in\mathbb{C}[\mathbf{z}]$, of type~$A_k$, $D_k$ or $E_k$. For
$j\in \mathbb{N}$, let $HH^j$ (resp.\ $HH_j$) be the Hochschild
cohomology (resp.\ homology) space in degree $j$ of
$A:=\mathbb{C}[\mathbf{z}]\,/\,\langle f\rangle$, and let $\nabla
f$ be the gradient of $f$. Then $HH^0\simeq HH_0\simeq A$,
$HH^1\simeq A\,\oplus\,\mathbb{C}^k$ and $HH_1\simeq
A^2\,/\,(A\nabla f)$, and for all~$j\geq 2$,
$HH^j\simeq HH_j\simeq\mathbb{C}^k$.
\end{proposition}

\begin{proposition}
Let $\Gamma$ be a finite subgroup of $\mathbf{SL}_2\mathbb{C}$ and
$f\in\mathbb{C}[\mathbf{z}]$ such that
$\mathbb{C}[x,\,y]^\Gamma\simeq\mathbb{C}[\mathbf{z}]\,/\,\langle
f\rangle$. For $j\in \mathbb{N}$, let $HH^j$ (resp.~$HH_j$) be the
Hochschild cohomology (resp.~homology) space in degree~$j$ of
$A:=\mathbb{C}[\mathbf{z}]\,/\,\langle f\rangle$, and let $\nabla
f$ be the gradient of $f$. Then $HH^0\simeq HH_0\simeq A$,
$HH^1\simeq (\nabla f \wedge A^3)\,\oplus\,\mathbb{C}^\mu$ and
$HH_1\simeq \nabla f \wedge A^3$, $HH^2\simeq
A \oplus \mathbb{C}^\mu$ and $HH_2\simeq A^3\,/\,(\nabla f\wedge
A^3)$, and for all $j\geq 3$, $HH^j\simeq
HH_j\simeq\mathbb{C}^\mu$, where $\mu$ is the Milnor number
of~$\mathcal{X}_\Gamma$.
\end{proposition}

For explicit computations, we shall make use of, and
develop a method suggested by M.~Kontsevich in the appendix
 of \cite{FK07}.

We will f\/irst study the case of singular curves of the plane in
Section~\ref{sec3}: we will use this method to recover the result that
C.~Fronsdal proved by direct calculations. Then we will ref\/ine it
by determining the dimensions of the cohomology and homology
spaces by means
of multivariate division and Groebner bases.

Next, in Section~\ref{sec4}, we will consider the case of Klein surfaces
$\mathcal{X}_\Gamma$. For $j\in \mathbb{N}$, we denote by $HH^j$
the Hochschild cohomology space in degree $j$ of
$\mathcal{X}_\Gamma$. We will f\/irst prove that $HH^0$ identif\/ies
with the space of polynomial functions on the singular
surface~$\mathcal{X}_\Gamma$. We will then prove that~$HH^1$
and~$HH^2$ are inf\/inite-dimensional. We will also determine, for~$j$ greater or equal to~$3$, the dimension of $HH^j$, by showing
that it is equal to the Milnor number of the
surface~$\mathcal{X}_\Gamma$. Finally, we will compute the
Hochschild homology spaces.

In Section~\ref{sec1.3} we begin by recalling
important classical results about deformations.

\subsection{Hochschild homology and cohomology and deformations of algebras}\label{sec1.3}

  Consider an associative $\mathbb{C}$-algebra,
denoted by $A$. The Hochschild cohomological complex of~$A$~is
\[
\xymatrix{C^0(A) \ar@{->}[r]^{d^{(0)}} &
C^1(A) \ar@{->}[r]^{d^{(1)}}
 & C^2(A) \ar@{->}[r]^{d^{(2)}}
 & C^3(A) \ar@{->}[r]^{d^{(3)}}
 & C^4(A) \ar@{->}[r]^{d^{(4)}}
 & \dots ,}\]
where the space $C^p(A)$ of $p$-cochains is def\/ined by $C^p(A)=0$
for $p\in -\mathbb{N}^*$, $C^0(A)=A$, and for $p\in \mathbb{N}^*$,
$C^p(A)$ is the space of $\mathbb{C}$-linear maps from $A^{\otimes
p}$ to~$A$. The dif\/ferential $d=\bigoplus_{i=0}^\infty d^{(p)}$ is
given by
\begin{gather*}
  \forall\; f\in C^p(A),   \quad
d^{(p)} f(a_0,\dots, a_p) = a_0f(a_1,\dots,a_p) \\
\phantom{\forall\; f\in C^p(A),   \quad} {}-\sum_{i=0}^{p-1}(-1)^if(a_0,\dots, a_ia_{i+1},\dots,\,a_p)
+(-1)^{p}f(a_0,\dots, a_{p-1})a_p.
\end{gather*}
 We may write it in terms of
the Gerstenhaber bracket\footnote{Recall that for $F\in C^p(A)$
and $H\in C^q(A)$, the Gerstenhaber product is the element
$F\bullet H\in C^{p+q-1}(A)$ def\/ined by $F\bullet
H(a_1,\dots, a_{p+q-1})=\sum_{i=0}^{p-1}(-1)^{i(q+1)}F(a_1,\dots, a_i, H(a_{i+1},\dots, a_{i+q}), a_{i+q+1},\dots, a_{p+q-1})$,
and the Gerstenhaber bracket is $[F, H]_G:=F\bullet
H-(-1)^{(p-1)(q-1)}H\bullet F$. See for example \cite{G63}, and
\cite[page~38]{BCKT05}.} $[\cdot,\cdot]_G$ and of the product $\mu$
of $A$, as follows  \[d^{(p)}f=(-1)^{p+1}[\mu, f]_G.\] Then we
def\/ine the Hochschild cohomology of $A$ as the cohomology of the
Hochschild cohomological complex associated to $A$, i.e.\
$HH^0(A):=\textrm{Ker}\, d^{(0)}$ and
 for $p\in \mathbb{N}^*$, $HH^p(A):=\textrm{Ker}\, d^{(p)} /\, \textrm{Im}\, d^{{(p-1)}}$.

We denote by $\mathbb{C}[[\hbar]]$ (resp.~$A[[\hbar]]$) the algebra of formal power series in the parameter
$\hbar$, with coef\/f\/icients in $\mathbb{C}$ (resp.~$A$). A
deformation of the map $\mu$ is a map $m$ from $A[[\hbar]]\times
A[[\hbar]]$ to $A[[\hbar]]$ which is
$\mathbb{C}[[\hbar]]$-bilinear and such that
\begin{gather*} \forall\; (s,\,t)\in
A[[\hbar]]^2,\qquad m(s,\,t)= st \mod \hbar A[[\hbar]], \\
\forall\; (s,\,t,\,u)\in
A[[\hbar]]^3,\qquad m(s,\,m(t,\,u))=m(m(s,\,t),\,u).
\end{gather*}
This means that there exists a sequence of bilinear maps $m_j$
from $A\times A$ to $A$ of which the f\/irst term $m_0$ is the
product of $A$ and such that
\begin{gather*}
\forall\; (a,\,b)\in A^2,\qquad
m(a,\,b)=\sum_{j=0}^{\infty}m_j(a,\,b)\hbar^j, \\
\forall\; n\in \mathbb{N},\qquad \sum_{i+j=n}m_i(a,m_j(b,c))=\sum_{i+j=n}m_i(m_j(a,b),c),\\
 \textrm{that is to say}\quad
  \sum_{i+j=n}[m_i,m_j]_G=0.
\end{gather*}
We say that $(A[[\hbar]],m)$ is a deformation
of the algebra $(A,\mu)$.
We say that the deformation is of order $p$ if the previous
formulae are satisf\/ied (only) for $n\leq p$.

The Hochschild cohomology plays an important
role in the study of deformations of the algebra $A$, by helping
us to classify them. In fact, if $\pi\in C^2(A)$, we may construct
a f\/irst order deformation $m$ of $A$ such that $m_1=\pi$ if and
only if $\pi\in \textrm{Ker}\,d^{(2)}$. Moreover, two f\/irst order
deformations are equivalent\footnote{Two deformations
$ m=\sum_{j=0}^{p}m_j\,\hbar^j,\ m_j\in C^2(A) $ and
$ m'=\sum_{j=0}^{p}m'_j\,\hbar^j,\ m'_j\in C^2(A) $
are called equivalent if there exists a sequence of linear maps
$\varphi_j$ from $A$ to $A$ of which the f\/irst term $\varphi_0$ is
the identity of $A$ and such that \begin{gather*}
 \forall\; a\in A,\qquad
\varphi(a)=\sum_{j=0}^{\infty}\varphi_j(a)\hbar^j, \\
 \forall\; n\in \mathbb{N},\qquad \sum_{i+j=n}\varphi_i(m_j(a, b))=\sum_{i+j+k=n}m'_i(\varphi_j(a), \varphi_k(b)).
\end{gather*}} if and only if their
dif\/ference is an element of $\textrm{Im}\,d^{(1)}$. So the set of
equivalence classes of f\/irst order deformations is
in bijection with $HH^2(A)$.

If $ m=\sum_{j=0}^{p}m_j \hbar^j$, $m_j\in C^2(A) $
is a deformation of order $p$, then we may extend $m$ to a~deformation of order $p+1$ if and only if there exists $m_{p+1}$
such that
\begin{gather*}
\forall\; (a,b,c)\in A^3,\qquad \sum_{i=1}^p\left(m_i(a,m_{p+1-i}(b,c))-m_i(m_{p+1-i}(a,b),c)\right)=-d^{(2)}m_{p+1}(a,b,c), \\
\textrm{i.e.}\ \  \sum_{i=1}^p[m_i,\,m_{p+1-i}]_G=2\,d^{(2)}m_{p+1}.
\end{gather*}
According to the graded Jacobi identity for
$[\cdot,\cdot]_G$, the last sum belongs to
$\textrm{Ker}\,d^{(3)}$. So~$HH^3(A)$ contains the obstructions
to extend a deformation of order $p$ to a deformation of order $p+1$.

The Hochschild homological complex of $A$ is
\[\xymatrix{\dots \ar@{->}[r]^{d_5} & C_4(A)\ar@{->}[r]^{d_4} &
C_3(A) \ar@{->}[r]^{d_3}
 & C_2(A) \ar@{->}[r]^{d_2}
 & C_1(A) \ar@{->}[r]^{d_1}
 & C_0(A),}\]
where the space of $p$-chains is given by $C_p(A)=0$ for $p\in
-\mathbb{N}^*$, $C_0(A)=A$, and for~$p\in \mathbb{N}^*$,
$C_p(A)=A\otimes A^{\otimes p}$. The dif\/ferential
$d=\bigoplus_{i=0}^\infty d_p$ is given by
\begin{gather*}
  d_p\,(a_0\otimes a_1\otimes\dots\otimes a_p) =a_0a_1\otimes
a_2\otimes\dots\otimes a_p \\
\qquad{}+\sum_{i=1}^{p-1}(-1)^i a_0\otimes
a_1\otimes\dots \otimes a_i a_{i+1}\otimes\dots \otimes a_p
+(-1)^p a_p a_0\otimes a_1\otimes\dots \otimes a_{p-1}.
\end{gather*} We
def\/ine the Hochschild homology of $A$ as the homology of the
Hochschild homological complex associated to $A$, i.e.\
$HH_0(A):=A\,/\,\textrm{Im}\ d_1$ and
 for $p\in \mathbb{N}^*$, $HH_p(A):=\textrm{Ker}\, d_p\, /\, \textrm{Im}\, d_{p+1}$.

\section{Presentation of the Koszul complex}\label{sec2}

\noindent We recall in this section some results about the Koszul
complex used below (see the appendix of~\cite{FK07}).

\subsection{Kontsevich theorem and notations}\label{sec2.1}

As in Section~\ref{sec1.2}, we consider
$R=\mathbb{C}[\mathbf{z}]$ and $(f_1,\dots,f_m)\in R^m$, and
we denote by $A$ the quotient $R\, /\, \langle f_1,\dots,
f_m\rangle$. We assume that we have a \emph{complete
intersection}, i.e.\ the dimension of the set of solutions of the
system
 $\{f_1(\mathbf{z})=\dots=f_m(\mathbf{z})=0\}$ is $n-m$.

We consider the dif\/ferential graded algebra
\[
\widetilde{T}=A[\eta_1,\dots,\eta_n;b_1,\dots,\,b_m]=
\frac{\mathbb{C}[z_1,\dots, z_n]}{\langle f_1,\dots,
f_m\rangle}[\eta_1,\dots,\eta_n;b_1,\dots,b_m],
\] where
$\eta_i:=\frac{\partial}{\partial z_i}$ is an odd variable (i.e.\
the $\eta_i$'s anticommute), and $b_j$ an even
variable (i.e.\ the $b_j$'s commute).

$\widetilde{T}$ is endowed with the dif\/ferential
\[
{d_{\widetilde{T}}=\sum_{j=1}^n \sum_{i=1}^m
\frac{\partial f_i}{\partial z_j}b_i\frac{\partial}{\partial
\eta_j}} ,
\]
and the Hodge grading, def\/ined by $\deg(z_i)=0$,
$\deg(\eta_i)=1$, $\deg(b_j)=2.$

We may now state the main theorem which will allow us to
calculate the Hochschild cohomology:

\begin{theorem}[Kontsevich]\label{kont}
Under the previous assumptions, the Hochschild cohomology of $A$
is isomorphic to the cohomology of the complex $(\widetilde{T},
d_{\widetilde{T}})$ associated with the differential graded algebra~$\widetilde{T}$.
\end{theorem}

\begin{remark}
Theorem~\ref{kont} may be seen as a generalization of the
Hochschild--Kostant--Rosenberg theorem to the case of
non-smooth spaces.
\end{remark}

There is no element of negative degree. So the
complex is as follows
\[\xymatrix{\widetilde{T}(0) \ar@{->}[r]^{\widetilde{0}} &
\widetilde{T}(1) \ar@{->}[r]^{d_{\widetilde{T}}^{(1)}}
 & \widetilde{T}(2) \ar@{->}[r]^{d_{\widetilde{T}}^{(2)}}
 & \widetilde{T}(3) \ar@{->}[r]^{d_{\widetilde{T}}^{(3)}}
 & \widetilde{T}(4) \ar@{->}[r]^{d_{\widetilde{T}}^{(4)}}
 & \dots }.\]
For each degree $p$, we choose a basis $\mathcal{B}_p$
of $\widetilde{T}(p)$. For example for $p=0,\dots, 3$, we may take
\begin{gather*}
\widetilde{T}(0)=A,\\
\widetilde{T}(1)=A\eta_1\oplus\dots\oplus A\eta_n, \\
\widetilde{T}(2)=A b_1\oplus\dots\oplus A b_m\oplus \bigoplus_{i<j}A\,\eta_i \eta_j, \\
\widetilde{T}(3)=\bigoplus_{\substack{i=1,\dots, m\\ j=1,\dots, n}}A\, b_i \eta_j\oplus\bigoplus_{i<j<k}A\,\eta_i \eta_j \eta_k.
\end{gather*}
Below we shall make use of the explicit matrices
${\rm Mat}_{\mathcal{B}_p,\mathcal{B}_{p+1}}(d_{\widetilde{T}}^{(p)})$.

 Set $H^0:=A$,
$H^1:=\textrm{Ker}\,d_{\widetilde{T}}^{(1)}$ and for
$j\geq 2$, $H^p:=\textrm{Ker}\,d_{\widetilde{T}}^{(p)}\,/\,\textrm{Im}\,d_{\widetilde{T}}^{(p-1)}$.
According to Theo\-rem~\ref{kont}, we have, for $p\in \mathbb{N}$,
$HH^p(A)\simeq H^p$.

There is an analogous of Theorem \ref{kont}
for the Hochschild homology. We consider the complex
\[
\widetilde{\Omega}=A[\xi_1,\dots, \xi_n; a_1,\dots, a_m],
\]
where $\xi_i$ is an odd variable and $a_j$ an even variable.
$\widetilde{\Omega}$ is endowed with the dif\/ferential
\[
d_{\widetilde{\Omega}}=\sum_{i=1}^n
\sum_{j=1}^m \frac{\partial f_j}{\partial
z_i}\xi_i\frac{\partial}{\partial a_j}  ,
\] and the Hodge
grading,
def\/ined by $\deg(z_i)=0$, $\deg(\xi_i)=-1$, $\deg(a_j)=-2.$

\begin{theorem}[Kontsevich]\label{konthom}
Under the previous assumptions, the Hochschild homology of $A$ is
isomorphic to the cohomology of the complex $(\widetilde{\Omega},
d_{\widetilde{\Omega}})$
\[\xymatrix{\dots \ar@{->}[r]^{d_{\widetilde{\Omega}}^{(-5)}} &
\widetilde{\Omega}(-4) \ar@{->}[r]^{d_{\widetilde{\Omega}}^{(-4)}}
 & \widetilde{\Omega}(-3) \ar@{->}[r]^{d_{\widetilde{\Omega}}^{(-3)}}
 & \widetilde{\Omega}(-2) \ar@{->}[r]^{d_{\widetilde{\Omega}}^{(-2)}}
 & \widetilde{\Omega}(-1) \ar@{->}[r]^{d_{\widetilde{\Omega}}^{(-1)}}
 & \widetilde{\Omega}(0) }.\]
\end{theorem}

 For each degree $p$, we will choose a basis
$\mathcal{V}_p$ of $\widetilde{\Omega}(p)$ and we will make use of
the explicit matrices ${\rm Mat}_{\mathcal{V}_p,\mathcal{V}_{p+1}}(d_{\widetilde{\Omega}}^{(p)})$.
Set $L^0:=A\,/\,\textrm{Im}\,d_{\widetilde{\Omega}}^{(-1)}$, and
for
$p\geq 1$, $L^{-p}:=\textrm{Ker}\,d_{\widetilde{\Omega}}^{(-p)}/\,\textrm{Im}\,d_{\widetilde{\Omega}}^{(-p-1)}$.
According to Theorem~\ref{konthom}, we have, for $p\in
\mathbb{N}$,
$HH_p(A)\simeq L^{-p}$.

For each ideal $J$ of
$\mathbb{C}[\mathbf{z}]$, we denote by $J_A$ the image of $J$ by
the canonical projection
\[
\mathbb{C}[\mathbf{z}]\rightarrow A=\mathbb{C}[\mathbf{z}]/\langle
f_1,\dots,f_m\rangle.
\] Similarly if $(g_1,\dots,g_r)\in A^r$
we denote by $\langle g_1,\dots,g_r\rangle _A$ the ideal of $A$
generated by
$(g_1,\dots,g_r)$.
Besides, if $g\in \mathbb{C}[\mathbf{z}]$, and if $J$ is an ideal
of $\mathbb{C}[\mathbf{z}]$, we set \[{\rm Ann}_J(g):=\{h\in
\mathbb{C}[\mathbf{z}]\, /\, hg=0\mod J\}.
\] In particular, $g$
does not divide $0$ in $\mathbb{C}[\mathbf{z}]\,/\,J$ if and only
if ${\rm Ann}_J(g)=J$. Finally, we denote by~$\nabla g$ the
gradient of a polynomial $g\in \mathbb{C}[\mathbf{z}]$.

From now on, we consider the case $m=1$ and
set $f:=f_1$. Moreover, we use the notation $\partial_j$ for the
partial derivative with respect to~$z_j$.

\subsection[Particular case where $n=1$ and $m=1$]{Particular case where $\boldsymbol{n=1}$ and $\boldsymbol{m=1}$}\label{sec2.2}

In the case where $n=1$ and $m=1$, according
to what we have seen, we have for $p\in \mathbb{N}$,
\[
\widetilde{T}(2p)=A b_1^p,\qquad \widetilde{T}(2p+1)=A
b_1^p \eta_1,\qquad \widetilde{\Omega}(-2p)=A a_1^p,\qquad
\widetilde{\Omega}(-2p-1)=A a_1^p \xi_1.
\]

We deduce
\[
H^0=L^0=A, \qquad H^1=\{g\eta_1\, /\, g\in A \mbox{and} \ g\partial_{1}f=0\},
\]
and for $p\in \mathbb{N}^*$,
\[
H^{2p}=\frac{Ab_1^p}{\{g
(\partial_{1}f) b_1^p\, /\, g\in A\}},\qquad \mbox{and}\qquad
 H^{2p+1}=\{gb_1^p\eta_1 \, /\, g\in A\ \textmd{and}\ g\partial_{1}f=0\}.
 \]

Similarly, for $p\in \mathbb{N}^*$,
\[
L^{-2p}=\{g a_1^p\, /\, g\in A\ \textrm{and}\
g\partial_{1}f=0\},
\]
 and for $p\in \mathbb{N}$,
\[
L^{-2p-1}=\frac{Aa_1^p\xi_1}{\{g(\partial_1f)a_1^p\xi_1\, /\, g\in A\}}.
\]

Now if $f=z_1^k$, then
\begin{gather*}
H^0=L^0=A=\mathbb{C}[z_1]\, /\, \langle z_1^k\rangle\simeq \mathbb{C}^{k},\\
H^1=\{g\eta_1\, /\, g\in A\ \textmd{and}\ kgz_1^{k-1}=0\}\simeq \mathbb{C}^{k-1},\qquad L^{-1}=\frac{A\xi_1}{\{g(kz_1^{k-1})\xi_1\, /\, g\in A\}}\simeq \mathbb{C}^{k-1},
\end{gather*}
and for $p\in \mathbb{N}^*$,
\[
H^{2p}\simeq
L^{-2p-1}\simeq\frac{Ab_1^p}{\{g (kz_1^{k-1}) b_1^p\, /\, g\in
A\}}\simeq \mathbb{C}^{k-1},
\] and for $p\in \mathbb{N}^*$,
\[
H^{2p+1}\simeq L^{-2p}\simeq\{g b_1^p\eta_1 \, /\, g\in A\ \textmd{and}\ kgz_1^{k-1}=0\}\simeq \mathbb{C}^{k-1}.
\]
See \cite{L98} for a similar calculation.

\section[Case $n=2$, $m=1$. Singular curves of the plane]{Case $\boldsymbol{n=2}$, $\boldsymbol{m=1}$. Singular curves of the plane}\label{sec3}

\subsection{Singular curves of the plane}\label{sec3.1}

\noindent In this section, we recall a result about the
weighted homogeneous functions, given in \cite[page~181]{AVGZ86}.

\begin{theorem}[Classif\/ication of weighted homogeneous functions, \cite{AVGZ86}]\label{Classification homogeneous functions}
The weighted homogeneous functions of two variables and of
modality zero reduce, up to equivalence, to the following list of
normal forms
$$\begin{array}{|l||c|c|c|c|c|}\hline
  \textrm{Type} & A_k & D_k & E_6 & E_7 & E_8  \\ \hline
  \textrm{Normal form} & z_1^{k+1}+z_2^2 & z_1^2z_2+z_2^{k-1} & z_1^3+z_2^4 & z_1^3+z_1z_2^3 & z_1^3+z_2^5
  \\\hline
\end{array}$$
\end{theorem}

The singularities of types $A_k$, $D_k$, $E_6$, $E_7$,
$E_8$ are called simple singularities.
In the two following sections, we will study the Hochschild
cohomology of $\mathbb{C}[\mathbf{z}]\,/\,\langle f\rangle$, where
$f$ is one of the normal forms of the preceding table.

\subsection{Description of the cohomology spaces}\label{sec3.2}

With the help of Theorem~\ref{kont} we calculate the
Hochschild cohomology of $A:=\mathbb{C}[z_1,z_2]\,/\,\langle
f\rangle$, where $f\in\mathbb{C}[z_1,z_2]$. We begin by
making cochains and dif\/ferentials explicit, by using the notations of Section~\ref{sec2.1}.

The various spaces of the complex are given by
\begin{alignat*}{3}
&  \widetilde{T}(0)=A,\qquad && \widetilde{T}(5)=A b_1^2\eta_1\oplus A b_1^2\eta_2, & \\
&  \widetilde{T}(1)=A\eta_1\oplus A\eta_2,\qquad & & \widetilde{T}(6)=A b_1^3\oplus A b_1^2\eta_1\eta_2, & \\
&  \widetilde{T}(2)=A b_1\oplus A\eta_1\eta_2,\qquad && \widetilde{T}(7)=A b_1^3\eta_1\oplus A b_1^3\eta_2, & \\
&  \widetilde{T}(3)=A b_1\eta_1\oplus A b_1\eta_2,\qquad && \widetilde{T}(8)=A b_1^4\oplus A b_1^3\eta_1\eta_2, & \\
&  \widetilde{T}(4)=A b_1^2\oplus A b_1\eta_1\eta_2,\qquad && \widetilde{T}(9)=A b_1^4 \eta_1\oplus A b_1^4\eta_2, &
\end{alignat*}
i.e., for an arbitrary $p\in \mathbb{N}^*$,
\[
\widetilde{T}(2p)=A b_1^p\oplus A
b_1^{p-1}\eta_1\eta_2,
\]
and for an arbitrary $p\in \mathbb{N}$,
\[
\widetilde{T}(2p+1)=A b_1^p
\eta_1\oplus A b_1^p\eta_2.
\]

As in \cite{FK07}, we denote by
$\frac{\partial}{\partial\eta_k}$ the partial derivative with
respect to the variable $\eta_k$, for $k\in\{1,2\}$. So,
for~$\{k,l\}=\{1,2\}$, we have
\[
\frac{\partial}{\partial\eta_k}(\eta_k\wedge\eta_l)=1\wedge\eta_l=-\eta_l\wedge1,
\]
hence
\[
d_{\widetilde{T}}^{(2)}(\eta_k\eta_l)=-\frac{\partial
f}{\partial z_k} b_1\eta_l+\frac{\partial
f}{\partial z_l} b_1\eta_k.
\]
The matrices of $d_{\widetilde{T}}$ are therefore given
by
\begin{gather*}
 {\rm Mat}_{\mathcal{B}_{2p},\mathcal{B}_{2p+1}}\big(d_{\widetilde{T}}^{(2p)}\big)=\left(
\begin{array}{cc}
  0 & \partial_{2}f \\
  0 & -\partial_{1}f
\end{array}
\right), \\
  {\rm Mat}_{\mathcal{B}_{2p+1},\mathcal{B}_{2p+2}}\big(d_{\widetilde{T}}^{(2p+1)}\big)=\left(
\begin{array}{cc}
  \partial_{1}f & \partial_{2}f \\
  0 & 0
\end{array}
\right).
\end{gather*}

We deduce a simpler expression for the cohomology spaces
\begin{gather*}
  H^0  =  A, \\
 H^1  =  \{g_1\eta_1+g_2\eta_2\, /\, (g_1,g_2)\in A^2\ \textmd{and}\
g_1 \partial_{1}f+g_2\partial_{2}f=0\}\\
\phantom{H^1}{} \simeq  \left\{\mathbf{g}=\left(
\begin{array}{c}
  g_1 \\
  g_2
\end{array}
\right)\in A^2\, /\, \mathbf{g}\cdot\nabla f=0\right\}.
\end{gather*}

For $p\in \mathbb{N}^*$,
\begin{gather*} H^{2p}  =  \frac{\{g_1 b_1^p+g_2
b_1^{p-1}\eta_1\eta_2\, /\, (g_1,\,g_2)\in A^2\ \textmd{and}\
g_2\partial_{1}f=g_2\partial_{2}f=0\}}{\{(g_1\partial_{1}f+g_2\partial_{2}f)b_1^p\,
/\, (g_1,g_2)\in A^2\}}\\
\phantom{H^{2p}}{} \simeq\frac{\left\{\mathbf{g}=\left(
\begin{array}{c}
  g_1 \\
  g_2
\end{array}
\right)\in A^2\, /\,
g_2 \partial_{1}f=g_2 \partial_{2}f=0\right\}}{\left\{\left(
\begin{array}{c}
  \mathbf{g}\cdot\nabla f \\
  0
\end{array}
\right)\, /\, \mathbf{g}\in A^2\right\}} \\
\phantom{H^{2p}}{} \simeq  \frac{A}{\langle \partial_{1}f, \partial_{2}f\rangle_A}\oplus \{g\in A\, /\, g\,\partial_{1}f=g\partial_{2}f=0\},\\
H^{2p+1}=\frac{\{g_1 b_1^p\eta_1+g_2 b_1^p\eta_2\, /\,
(g_1,g_2)\in A^2\ \textmd{and}\
g_1\partial_{1}f+g_2\partial_{2}f=0\}}{\{g_2(\partial_{2}fb_1^p\eta_1-\partial_{1}fb_1^p\eta_2)\,
/\, g_2\in A\}}\\
\phantom{H^{2p+1}}{} \simeq\frac{\left\{\mathbf{g}=\left(
\begin{array}{c}
  g_1 \\
  g_2
\end{array}
\right)\in A^2\, /\, \mathbf{g}\cdot\nabla
f=0\right\}}{\left\{g_2\left(
\begin{array}{c}
  \partial_{2}f \\
  -\partial_{1}f
\end{array}
\right)\, /\, g_2\in A\right\}}.
\end{gather*}

\begin{remark}\label{resFronsdal}
We recover a result of \cite{FK07} (here, we use the notations of
\cite{FK07}). According to Theo\-rem~3.8 of \cite{FK07}, we have
${\rm Hoch}_{2p}={\rm Hoch}_{2p,p}\oplus {\rm Hoch}_{2p,p+1}$ and
${\rm Hoch}_{2p+1}={\rm Hoch}_{2p+1,p+1}$, so ${\rm Hoch}_{2p,k}=0$ if $k\notin
\{p, p+1\}$,
and $Hoch_{2p+1,k}=0$ if $k\neq p+1$.
By using Section 4.1 of~\cite{FK07}, we deduce $H^{2p,k}=0$ if
$k\notin \{p,p+1\}$, and $H^{2p+1,k}=0$ if $k\neq
p+1$.
Hence $H^{2p}=H^{2p,p}\oplus H^{2p,p+1}$ and
$H^{2p+1}=H^{2p+1,p+1}$.
So Theorem 4.9 of \cite{FK07} gives the cohomology spaces which we
have just obtained.
\end{remark}

It remains to determine these spaces more explicitly. This will be done in the two following sections.

\subsection[Explicit calculations in the particular case where $f$ has separate variables]{Explicit calculations in the particular case where $\boldsymbol{f}$ has separate variables}\label{sec3.3}

In this section, we consider the polynomial
$f=a_1z_1^k+a_2z_2^l$, with $k\geq 2$, $l\geq 2$, and
$(a_1,a_2){\in}(\mathbb{C^*})^2$.
The partial derivatives of $f$ are
$\partial_1f=ka_1z_1^{k-1}$ and
$\partial_2f=la_2z_2^{l-1}$.

We already have
\[
H^0=\mathbb{C}[z_1,z_2]\,/\,\langle
a_1z_1^k+a_2z_2^l\rangle .
\]

 Besides, as $f$ is weighted homogeneous,
Euler's formula gives
$ \frac{1}{k}z_1\partial_1f+\frac{1}{l}z_2\partial_2f=f$.
So we have the inclusion $\langle f\rangle\subset\langle
\partial_1f, \partial_2f\rangle$, hence
\[\frac{A}
{\langle \partial_{1}f,  \partial_{2}f\rangle_A}\simeq
\frac{\mathbb{C}[z_1, z_2]}{\langle \partial_{1}f,
\partial_{2}f\rangle}\simeq {\rm Vect}\big(z_1^iz_2^j\, /\, i\in\llbr 0, k-2\rrbr,\
j\in\llbr 0,\,l-2\rrbr\big).\]
But $\partial_1f$ and $f$ are relatively prime, just as
$\partial_2f$ and $f$ are, hence if $g\in A$ satisf\/ies
$g\partial_1f=0 \mod \langle f\rangle$, then $g\in \langle
f\rangle$, i.e. $g$ is zero in $A$.
So, \[
H^{2p}\simeq {\rm Vect}\big(z_1^iz_2^j\, /\, i\in\llbr
0, k-2\rrbr,\
j\in\llbr 0, l-2\rrbr\big)\simeq \mathbb{C}^{(k-1)(l-1)} .\]

 We now determine the set
\[
\left\{\mathbf{g}=\left(
\begin{array}{c}
  g_1 \\
  g_2
\end{array}
\right)\in A^2\, /\, \mathbf{g}\cdot\nabla f=0\right\}.
\]
First we have
\[
\langle f, \partial_1f\rangle=\langle
a_1z_1^k+a_2z_2^l, z_1^{k-1}\rangle=\langle
z_2^l, z_1^{k-1}\rangle.
\] So the only monomials which are not in
this ideal are the elements $z_1^iz_2^j$ with $i\in\llbr
0,k-2\rrbr$ and $j\in\llbr 0,l-1\rrbr$.
Every polynomial $P\in \mathbb{C}[\mathbf{z}]$ may be written in
the form
\[ P=\alpha f+\beta
\partial_1f+\sum_{\substack{i=0,\dots, k-2 \\ j=0,\dots,
l-1}}a_{ij}z_1^iz_2^j ,\]
with $\alpha, \beta\in \mathbb{C}[\mathbf{z}]$ and $a_{ij}\in
\mathbb{C}$.
Therefore, the polynomials $P\in \mathbb{C}[\mathbf{z}]$ such that
$P\partial_2f\in \langle f, \partial_1f\rangle$ are the elements
\[ P=\alpha f+\beta
\partial_1f+\sum_{\substack{i=0,\dots, k-2 \\ j=1,\dots,
l-1}}a_{ij}z_1^iz_2^j.\]

So we have calculated ${\rm Ann}_{\langle f, \partial_1f\rangle}(\partial_2f)$.
Let $\mathbf{g}=\left(%
\begin{array}{c}
  g_1 \\
  g_2
\end{array}
\right)\in A^2$ satisfy the equation
\begin{gather}\label{21eq1}
\mathbf{g}\cdot\nabla f=0 \mod \langle f\rangle.
\end{gather}
Then we have
\begin{gather*}
g_2\partial_2f=0 \mod \langle f,\partial_1f\rangle,
\end{gather*}
i.e.\ $g_2\in {\rm Ann}_{\langle f, \partial_1f\rangle}(\partial_2f)$,
i.e.\ again
\begin{gather*}
g_2=\alpha f+\beta\partial_1f+\sum_{\substack{i=0,\dots, k-2
\\ j=1,\dots,
l-1}}a_{ij}z_1^iz_2^j,\qquad \textrm{with}\  \ (\alpha,\,\beta)\in
\mathbb{C}[\mathbf{z}]^2.
\end{gather*}
It follows that
\begin{gather*}
g_1\partial_1f+\alpha
f\partial_2f+\beta\partial_1f\partial_2f+\sum_{\substack{i=0,\dots,
k-2 \\ j=1,\dots, l-1}}a_{ij}z_1^iz_2^j \partial_2f\in \langle
f\rangle.
\end{gather*}
From the equality $z_2\partial_2f=lf-\frac{l}{k}z_1\partial_1f$,
one deduces
\begin{gather*}
\partial_1f\left(g_1+\beta\partial_2f-\frac{l}{k}\sum_{\substack{i=0,\dots,
k-2 \\ j=1,\dots, l-1}}a_{ij}z_1^{i+1}z_2^{j-1}\right)\in \langle
f\rangle,
\end{gather*}
i.e.
\[g_1=-\beta\partial_2f+\frac{l}{k}\sum_{\substack{i=0,\dots, k-2 \\
j=1,\dots, l-1}}a_{ij}z_1^{i+1}z_2^{j-1}+\delta f,\qquad \textrm{with}\ \
\delta\in \mathbb{C}[\mathbf{z}].
\]  Then we verify that the
elements $g_1$ and $g_2$ obtained in this
way are indeed solutions of equation~(\ref{21eq1}).

Finally, we have
\begin{gather*}
\left\{\mathbf{g}\in A^2\, /\,
\mathbf{g}\cdot\nabla f=0 \right\}\\
\qquad{}=\left\{ -\beta\left(
\begin{array}{c}
  \partial_2f \\
  -\partial_1f
\end{array}
\right)+\sum_{\substack{i=0,\dots, k-2 \\ j=1,\dots,
l-1}}a_{ij}z_1^{i}z_2^{j-1}\left(
\begin{array}{c}
  \frac{l}{k}z_1 \\
  z_2
\end{array}
\right) \Big/\, \beta\in A\ \textrm{and}\ a_{ij}\in
\mathbb{C}\right\}.
\end{gather*}
We immediately deduce the cohomology spaces of odd degree:
\begin{alignat*}{3}
& \forall\; p\geq 1,\quad &&  H^{2p+1}  \simeq  \mathbb{C}^{(k-1)(l-1)},& \\
 &&& H^{1}  \simeq  \mathbb{C}^{(k-1)(l-1)}\oplus \mathbb{C}[z_1,\,z_2]/\langle a_1z_1^k+a_2z_2^l \rangle,
\end{alignat*}
where the direct sum results from the following argument: if we have
\[
-\beta\left(
\begin{array}{c}
  \partial_2f \\
  -\partial_1f
\end{array}%
\right)=\sum_{\substack{i=0,\dots, k-2 \\ j=1,\dots,
l-1}}a_{ij}z_1^{i}z_2^{j-1}\left(
\begin{array}{c}
  \frac{l}{k}z_1 \\
  z_2
\end{array}%
\right)\mod \langle f\rangle,
\]
 then
\[
v:= -\beta l
a_2 z_2^{l-1}-\sum_{\substack{i=0,\dots, k-2 \\ j=1,\dots,
l-1}}\frac{l}{k} a_{ij}z_1^{i+1}z_2^{j-1}\in \langle
f\rangle .\]
And by a Euclidian division in
$\left(\mathbb{C}[z_2]\right)[z_1]$, we may write $\beta=f q+r$,
where the $z_1$-degree of $r$ is smaller or equal to $k-1$. So the
$z_1$-degree of $v$ is also smaller or equal to $k-1$, thus
$v\in \langle f\rangle$ implies $\beta=0$ and~$a_{ij}=0$.

\begin{remark}
We obtain in particular the cohomology for the cases where
$f=z_1^{k+1}+z_2^2$ ($k\in \mathbb{N}^*$), $f=z_1^3+z_2^4$ and
$f=z_1^3+z_2^5$. These cases correspond respectively to the
weighted homogeneous
functions of types $A_k$, $E_6$ and $E_8$ given in Theorem~\ref{Classification homogeneous functions}.
\end{remark}

The table below summarizes the results we have just
obtained for the three particular cases
$$\begin{array}{|l||l|l|l|}
  \hline
\tsep{0.5mm}    & H^0 & H^1 & H^{p},\ p\geq 2 \\
  \hline\hline
\tsep{1mm}\bsep{1mm}  A_k & \mathbb{C}[\mathbf{z}]\, /\, \langle z_1^{k+1}+z_2^2\rangle & \mathbb{C}[\mathbf{z}]\, /\, \langle z_1^{k+1}+z_2^2\rangle\oplus \mathbb{C}^k & \mathbb{C}^k
  \\ \hline
\tsep{1mm}\bsep{1mm}  E_6 & \mathbb{C}[\mathbf{z}]\, /\, \langle z_1^3+z_2^4\rangle & \mathbb{C}[\mathbf{z}]\, /\, \langle z_1^3+z_2^4\rangle\oplus \mathbb{C}^6 & \mathbb{C}^6
  \\ \hline
\tsep{1mm}\bsep{1mm}  E_8 & \mathbb{C}[\mathbf{z}]\, /\, \langle z_1^3+z_2^5\rangle & \mathbb{C}[\mathbf{z}]\, /\, \langle z_1^3+z_2^5\rangle\oplus \mathbb{C}^8 & \mathbb{C}^8  \\
  \hline
\end{array}$$

The cases where $f=z_1^2z_2+z_2^{k-1}$ and
$f=z_1^3+z_1z_2^3$, i.e.\ respectively $D_k$ and $E_7$, will be
studied in the next section.

\subsection[Explicit calculations for $D_k$ and $E_7$]{Explicit calculations for $\boldsymbol{D_k}$ and $\boldsymbol{E_7}$}\label{sec3.4}

 To study these particular cases, we use the following
result about Groebner bases (Theorem~\ref{Macaulay}). First,
recall the def\/inition of a Groebner basis. For $g\in
\mathbb{C}[\mathbf{z}]$, we denote by ${\rm lt}(g)$ its leading
term (for the lexicographic order). Given a non-trivial ideal $J$
of $\mathbb{C}[\mathbf{z}]$, a \emph{Groebner basis} of $J$ is a
f\/inite subset $G_J$ of $J\,\backslash\,\{0\}$ such that for all
$f\in J\,\backslash\,\{0\}$, there exists $g\in G_J$ such that
${\rm lt}(g)$ divides~${\rm lt}(f)$. See \cite{RSP02} for more
details.

\begin{definition}
Let $J$ be a non-trivial ideal of $\mathbb{C}[\mathbf{z}]$ and let
$G_J:=[g_1,\dots,g_r]$ be a Groebner basis of $J$. We call
\emph{set of the $G_J$-standard terms}, the set of all monomials
of $\mathbb{C}[\mathbf{z}]$ that are not divisible by any of ${\rm
lt}(g_1),\dots,{\rm lt}(g_r)$.
\end{definition}

\begin{theorem}[Macaulay]\label{Macaulay}
The set of the $G_J$-standard terms forms a basis of the quotient
vector space $\mathbb{C}[\mathbf{z}]\, /\, J$.
\end{theorem}

\subsubsection[Case of $f=z_1^2z_2+z_2^{k-1}$, i.e. $D_k$]{Case of $\boldsymbol{f=z_1^2z_2+z_2^{k-1}}$, i.e.\ $\boldsymbol{D_k}$}\label{sec3.4.1}

Here we have
\[
f=z_1^2z_2+z_2^{k-1} ,
\qquad \partial_1f=2z_1z_2 \qquad \mbox{and}\qquad
\partial_2f=z_1^2+(k-1)z_2^{k-2} .
\]
A Groebner basis of the ideal $\langle f, \partial_2f\rangle$ is
\[
B:=[b_1, b_2]=[z_1^2+(k-1)z_2^{k-2}, z_2^{k-1}] .
\]
So the set of the standard terms is
\[
\{z_1^iz_2^j\, /\, i\in
\{0, 1\}\ \textrm{and}\ j\in \llbr 0,\ k-2\rrbr\} .
\]
We may now solve the equation $p \partial_1f=0$ in
$\mathbb{C}[\mathbf{z}]\, /\, \langle f, \partial_2f\rangle$.
In fact, by writing
\[
p:= \sum_{\substack{i=0,1 \\
j=0,\dots, k-2}}a_{ij}z_1^iz_2^j ,
\] the equation becomes \[ q:=\sum_{\substack{i=0,1 \\
j=0,\dots, k-2}}a_{ij}z_1^{i+1}z_2^{j+1}\in \langle
f, \partial_2f\rangle.\] We look for the normal form of the
element $q$ modulo the ideal $\langle
f, \partial_2f\rangle$.

The multivariate division of $q$ by $B$ is $q=q_1b_1+q_2b_2+r$
with $r=\sum_{j=0}^{k-3}a_{0,j}z_1z_2^{j+1}$.\\
Thus the solution in $\mathbb{C}[\mathbf{z}]\ /\ \langle
f,\,\partial_2f\rangle$ is
\[p=a_{0,k-2}z_2^{k-2}+\sum_{j=0}^{k-2}a_{1,j}z_1z_2^j.\]
But the equation
\begin{gather*}
\mathbf{g}\cdot\nabla f=0 \mod \langle f\rangle
\end{gather*}
yields
\begin{gather*}
g_1\partial_1f=0 \mod \langle f, \partial_2f\rangle,
\end{gather*}
i.e.
\begin{gather*}
g_1=\alpha
f+\beta\partial_2f+az_2^{k-2}+\sum_{j=0}^{k-2}b_jz_1z_2^j,\qquad
\textrm{with}  \ \ (\alpha,\beta)\in \mathbb{C}[\mathbf{z}]^2\ \
\textrm{and}\  \ a,b_j\in \mathbb{C}.
\end{gather*}
Hence
\begin{gather*}
g_2\partial_2f+\beta\partial_1f\,
\partial_2f+az_2^{k-2}\partial_1f+
\sum_{j=0}^{k-2}b_jz_1z_2^j\partial_1f\in \langle f\rangle.
\end{gather*}
And with the equalities,
\begin{gather*}
  z_2^{k-1}=\frac{1}{2-k}(f-z_2\partial_2f)=-\frac{1}{2-k}z_2\partial_2f\mod
\langle f\rangle,
\end{gather*}
and
\begin{gather*}
 \frac{k-2}{2}z_1\partial_1f+z_2\partial_2f=(k-1)f\quad
\textrm{(Euler)},
\end{gather*}
 we obtain
\begin{gather*}
\partial_2f\left(g_2+\beta\partial_1f -\frac{2a}{2-k}z_1z_2+
\sum_{j=0}^{k-2}b_j\frac{2}{2-k}z_2^{j+1}\right)\in \langle
f\rangle.
\end{gather*}
i.e.,
\[
g_2=-\beta\partial_1f+\frac{2a}{2-k}z_1z_2-
\sum_{j=0}^{k-2}b_j\frac{2}{2-k}z_2^{j+1}+\delta f,\qquad
\textrm{with}\ \ \delta\in \mathbb{C}[\mathbf{z}].
\]  So
\begin{gather*}
\left\{\mathbf{g}\in A^2\, /\,
\mathbf{g}\cdot\nabla f=0 \right\}\\
=\left\{ \beta\left(%
\begin{array}{c}
  \partial_2f \\
  -\partial_1f
\end{array}
\right)+a\left(
\begin{array}{c}
  z_2^{k-2} \\
  \frac{2}{2-k}z_1z_2 \\
\end{array}%
\right)+\sum_{j=0}^{k-2}b_jz_2^j\left(%
\begin{array}{c}
  z_1 \\
  -\frac{2}{2-k}z_2
\end{array}
\right) \Big/\,
  \beta\in A, \
  a,\,b_j\in \mathbb{C}
 \right\}.
\end{gather*}

 On the other hand, a Groebner basis of $\langle
\partial_1f, \partial_2f\rangle$ is
$[z_1^2+(k-1)z_2^{k-2}, z_1z_2, z_2^{k-1}]$,
thus
\[
\mathbb{C}[\mathbf{z}]\, /\, \langle
\partial_1f, \partial_2f\rangle\simeq {\rm Vect}\big(z_1, 1, z_2,\dots, z_2^{k-2}\big).
\]

Let us summarize (by using, for the direct sum, the same
argument as in Section \ref{sec3.2}):
\begin{gather*}
  H^0=\mathbb{C}[\mathbf{z}]\, /\, \langle z_1^2z_2+z_2^{k-1}\rangle,\\
H^1\simeq \mathbb{C}[\mathbf{z}]\, /\, \langle z_1^2z_2+z_2^{k-1}\rangle\oplus \mathbb{C}^k,\\
H^{2p}\simeq \mathbb{C}^k,\\
H^{2p+1}\simeq \mathbb{C}^k .
\end{gather*}

\subsubsection[Case of $f=z_1^3+z_1z_2^3$, i.e. $E_7$]{Case of $\boldsymbol{f=z_1^3+z_1z_2^3}$, i.e.\ $\boldsymbol{E_7}$}\label{sec3.4.2}

Here we have $\partial_1f=3z_1^2+z_2^3$ and
$\partial_2f=3z_1z_2^2$.
A Groebner basis of the ideal $\langle f, \partial_1f\rangle$ is
$[3z_1^2+z_2^3, z_1z_2^3, z_2^6]$, and a Groebner basis of
$\langle
\partial_1f, \partial_2f\rangle$ is
$[3z_1^2+z_2^3, z_1z_2^2, z_2^5]$.
By an analogous proof, we obtain
\begin{gather*}
H^0=\mathbb{C}[\mathbf{z}]\, /\, \langle z_1^3+z_1z_2^3\rangle,\\
H^1\simeq \mathbb{C}[\mathbf{z}]\, /\, \langle z_1^3+z_1z_2^3\rangle\oplus \mathbb{C}^7,\\
H^{2p}\simeq \mathbb{C}^7,\\
H^{2p+1}\simeq \mathbb{C}^7 .
\end{gather*}

\subsection{Homology}\label{sec3.5}

The study is the same as the one of the
Hochschild cohomology: to get the Hochschild homology is
equivalent to compute the cohomology of the complex
$(\widetilde{\Omega},\ d_{\widetilde{\Omega}})$ described in
Section~\ref{sec2.1}.
We have $\widetilde{\Omega}(0)=A$,
$\widetilde{\Omega}(-2p)=Aa_1^p\oplus Aa_1^{p-1}\xi_1\xi_2$ for
$p\in \mathbb{N}^*$, and
$\widetilde{\Omega}(-2p-1)=Aa_1^p\xi_1\oplus Aa_1^{p}\xi_2$ for
$p\in \mathbb{N}$. This def\/ines the bases $\mathcal{V}_{p}$. The
dif\/ferential is
$d_{\widetilde{\Omega}}=(\xi_1\partial_1f+\xi_2\partial_2f)\frac{\partial}{\partial
a_1}$.

 So we obtain, for $p\in \mathbb{N}^*$, the matrices
\[
{\rm Mat}_{\mathcal{V}_{-2p},\mathcal{V}_{-2p+1}}
\left(d_{\widetilde{\Omega}}^{(-2p)}\right)=\left(%
\begin{array}{cc}
  p\partial_1f & 0 \\
  p\partial_2f & 0
\end{array}%
\right),
\]
and
\[
{\rm Mat}_{\mathcal{V}_{-2p-1},\mathcal{V}_{-2p}}\left(d_{\widetilde{\Omega}}^{(-2p-1)}\right)=\left(%
\begin{array}{cc}
  0 & 0 \\
  -p\partial_2f & p\partial_1f
\end{array}%
\right).
\]

The cohomology spaces read as
\begin{gather*}
  L^0  =  A, \qquad
  L^{-1}  =  \frac{A^2}{\{g\nabla f\, /\, g\in A\}}.
\end{gather*}

For $p\in \mathbb{N}^*$,
\begin{gather*}
  L^{-2p}  \simeq  \frac{\left\{\left(%
\begin{array}{c}
  g_1 \\
  g_2
\end{array}
\right)\in A^2\, \Big/\, pg_1\partial_1f=pg_1\partial_2f=0\right\}}
{\left\{\left(\begin{array}{c}
  0 \\
  -(p+1)g_1\partial_2f+(p+1)g_2\partial_1f
\end{array}\right) \Big/ \left(%
\begin{array}{c}
  g_1 \\
  g_2
\end{array}%
\right)\in A^2\right\}}\\
\phantom{L^{-2p}}{} \simeq  \left\{g\in A\, /\, g\partial_1f=g\partial_2f=0\right\}\oplus \frac{A}{\langle\nabla f\rangle_A}.
\end{gather*}

For $p\in \mathbb{N}^*$,
\begin{gather*}
L^{-2p-1}  \simeq  \frac{\left\{\left(%
\begin{array}{c}
  g_1 \\
  g_2
\end{array}%
\right)\in A^2\, \Big/\, -pg_1\partial_2f+pg_2\partial_1f=0\right\}}
{\left\{\left(\begin{array}{c}
  (p+1)g_1\partial_1f \\
  (p+1)g_1\partial_2f
\end{array}\right) \Big/ \left(%
\begin{array}{c}
  g_1 \\
  g_2
\end{array}%
\right)\in A^2\right\}}\simeq\frac{\{\mathbf{g}\in A^2\, /\, \det(\nabla f,\,\mathbf{g})=0\}}  {\{g\nabla f\ /\ g\in A\}}.
\end{gather*}

From now on, we assume that $f$ has separate
variables, or $f$ is of type $D_k$ or $E_7$. Then we have
$\left\{g\in A\, /\, g\partial_1f=g\partial_2f=0\right\}=\{0\}$, and
according to Euler's formula, for $p\in \mathbb{N}$,
$L^{-2p}\simeq \frac{A}{\langle\nabla
f\rangle_A}\simeq\frac{\mathbb{C}[\mathbf{z}]}{\langle\nabla
f\rangle}$.
For the computation of $\{\mathbf{g}\in A^2\, /\, \det(\nabla
f, \mathbf{g})=0\}$ and $\frac{A^2}{\{g\nabla f\, /\, g\in A\}}$,
we proceed with Groebner bases as in Section~\ref{sec3.3}. For example, we do it for $f=z_1^2z_2+z_2^{k-1}$ (i.e.\ type $D_k$).

Let $\mathbf{g}\in A^2$ be such
that $\det(\nabla f,\mathbf{g})=0$. Then $g_2\partial_1f=0\mod
\langle f,\partial_2f\rangle$, i.e., according to Section~\ref{sec3.4.1},
\begin{gather*}
g_2=\alpha
f+\beta\partial_2f+az_2^{k-2}+\sum_{j=0}^{k-2}b_jz_1z_2^j,
\end{gather*}
with $(\alpha,\beta)\in \mathbb{C}[\mathbf{z}]^2$ and
$a,b_j\in \mathbb{C}$. Hence
\begin{gather*}
-g_1\partial_2f+\alpha
f\partial_1f+\beta\partial_2f\partial_1f+az_2^{k-2}\partial_1f+\sum_{j=0}^{k-2}b_jz_1z_2^j\partial_1f\in
\langle f\rangle.
\end{gather*}
With the equalities,
\begin{gather*}
  z_2^{k-1}=\frac{1}{2-k}(f-z_2\partial_2f)=-\frac{1}{2-k}z_2\partial_2f\mod
\langle f\rangle,
\end{gather*}
and
\begin{gather*}
 \frac{k-2}{2}z_1\partial_1f+z_2\partial_2f=(k-1)f\quad
\textrm{(Euler)},
\end{gather*}
 we obtain
\begin{gather*}
\partial_2f\left(-g_1+\beta\partial_1f -\frac{2a}{2-k}z_1z_2+
\sum_{j=0}^{k-2}b_j\frac{2}{2-k}z_2^{j+1}\right)\in \langle
f\rangle.
\end{gather*}
i.e.,
\begin{gather*}
 g_1=\beta\partial_1f-\frac{2a}{2-k}z_1z_2+
\sum_{j=0}^{k-2}b_j\frac{2}{2-k}z_2^{j+1}+\delta f,\qquad
\textrm{with}\  \ \delta\in \mathbb{C}[\mathbf{z}].
\end{gather*} So
\begin{gather*}
\left\{\mathbf{g}\in A^2\, /\,
\det(\nabla f,\,\mathbf{g})=0 \right\}\\
\qquad{}=\left\{ \beta\nabla f+a\left(%
\begin{array}{c}
  -\frac{2}{2-k}z_1z_2 \\
  z_2^{k-2}
\end{array}
\right)+\sum_{j=0}^{k-2}b_jz_2^j\left(%
\begin{array}{c}
  \frac{2}{2-k}z_2 \\
  z_1
\end{array}%
\right) \Big/\,
  \beta\in A, \
  a,b_j\in \mathbb{C}
 \right\}.
 \end{gather*}

We have $\{g\nabla f\, /\, g\in
A\}\subset \{\mathbf{g}\in A^2\, /\, \det(\nabla
f, \mathbf{g})=0\}$, thus
\[
\dim\left( A^2\,/\,\{g\nabla f\, /\,
g\in A\}\right) \geq \dim\left( A^2\,/\,\{\mathbf{g}\in A^2\, /\,
\det(\nabla f, \mathbf{g})=0\}\right).
\] Since
$A^2\,/\,\{\mathbf{g}\in A^2\, /\, \det(\nabla
f,\mathbf{g})=0\}\simeq \{\det(\nabla f,\mathbf{g})\, /\,
\mathbf{g}\in A^2\}$, and since the map
\[g\in A\mapsto
\det\left(\nabla f,\left(%
\begin{array}{c}
  g \\
  0
\end{array}%
\right)\right)\in \{\det(\nabla f,\mathbf{g})\, /\, \mathbf{g}\in
A^2\}
\] is injective, we deduce
that $A^2\,/\,\{g\nabla f\, /\, g\in A\}$ is inf\/inite-dimensional.

We collect in the following table the results
for the Hochschild homology in the various cases
$$
\begin{array}{|l||l|l|l|}\hline
 \tsep{0.5mm}\bsep{0.5mm} \textrm{Type} & HH_0=A & HH_1 & HH_p,\ p\geq 2  \\ \hline\hline
\tsep{0.5mm}\bsep{0.5mm} A_k & \mathbb{C}[\mathbf{z}]\,/\,\langle z_1^{k+1}+z_2^2\rangle &
A^2\,/\,A\nabla f & \mathbb{C}^{k} \\ \hline
\tsep{0.5mm}\bsep{0.5mm} D_k & \mathbb{C}[\mathbf{z}]\,/\,\langle z_1^2z_2+z_2^{k-1}\rangle
& A^2\,/\,A\nabla f & \mathbb{C}^{k} \\ \hline
\tsep{0.5mm}\bsep{0.5mm} E_6 & \mathbb{C}[\mathbf{z}]\,/\,\langle z_1^3+z_2^4\rangle &
A^2\,/\,A\nabla f & \mathbb{C}^{6} \\ \hline
\tsep{0.5mm}\bsep{0.5mm} E_7 & \mathbb{C}[\mathbf{z}]\,/\,\langle z_1^3+z_1z_2^3\rangle &
A^2\,/\,A\nabla f & \mathbb{C}^{7} \\ \hline
\tsep{0.5mm}\bsep{0.5mm} E_8 & \mathbb{C}[\mathbf{z}]\,/\,\langle z_1^3+z_2^5\rangle &
A^2\,/\,A\nabla f & \mathbb{C}^{8} \\ \hline
\end{array}
$$

\section[Case $n=3$, $m=1$. Klein surfaces]{Case $\boldsymbol{n=3}$, $\boldsymbol{m=1}$. Klein surfaces}\label{sec4}

\subsection{Klein surfaces}\label{sec4.1}

Given a f\/inite group $G$ acting on $\mathbb{C}^n$, we
associate to it, according to Erlangen program of Klein, the
quotient space $\mathbb{C}^n/G$, i.e.\ the space whose points are
the orbits under the action of~$G$; it is an algebraic variety,
and the polynomial functions on this variety are the
polynomial functions on $\mathbb{C}^n$ which are $G$-invariant.
In the case of $\mathbf{SL}_2\mathbb{C}$, invariant theory allows
us to associate a~polynomial to any f\/inite subgroup, as explained
in Proposition~\ref{invariants}. Thus, to every f\/inite subgroup of
$\mathbf{SL}_2\mathbb{C}$ is associated the zero set of this
polynomial; it is an algebraic variety, called a Klein surface.

In this section we recall some results about these
surfaces.
 See the references \cite{S77} and \cite{CCK99} for more details.

\begin{proposition}
Every finite subgroup of $\mathbf{SL}_2\mathbb{C}$ is conjugate to one of the following groups:
\begin{itemize}\itemsep=0pt
\item $A_k$ (cyclic), $k\geq 1$, $|A_k|=k$;
\item $D_k$ (dihedral), $k\geq 1$, $|D_k|=4k$;
\item $E_6$ (tetrahedral), $|E_6|=24$;
\item $E_7$ (octahedral), $|E_7|=48$;
\item $E_8$ (icosahedral), $|E_8|=120$.
\end{itemize}
\end{proposition}

\begin{proposition}\label{invariants}
Let $G$ be one of the groups of the preceding list. The ring of
invariants is the following
\[\mathbb{C}[x,y]^G=\mathbb{C}[e_1,e_2,e_3]=\mathbb{C}[e_1,e_2]
\oplus e_3\mathbb{C}[e_1,e_2]\simeq
\mathbb{C}[z_1,z_2,z_3]\,/\,\langle f\rangle,\] where the
invariants $e_j$ are homogeneous polynomials, with $e_1$ and $e_2$
algebraically independent, and where $f$
is a weighted homogeneous polynomial with an isolated singularity at the origin.
These polynomials are given in the following table.
\end{proposition}

We call Klein surface the algebraic hyper-surface
def\/ined by $\{\mathbf{z}\in \mathbb{C}^3\, /\, f(\mathbf{z})=0\}.$
\vspace{1mm}

\begin{scriptsize}
\centerline{\begin{tabular}{|@{\,\,}l@{\,\,}||l|l|l@{}|} \hline
\tsep{0.5mm}\bsep{0.5mm} $G$ & $e_1$, $e_2$, $e_3$ & $f$ & $\mathbb{C}[z_1,z_2,z_3]\,/\,\langle\partial_1 f,\partial_2 f,\partial_3 f\rangle$  \\ \hline
\hline
$A_k$ &
 \begin{minipage}{4cm}
\tsep{1ex}$e_1=x^k$\\ $e_2=y^k$ \\$e_3=xy$ \bsep{1ex}
\end{minipage} & $-k(z_1z_2-z_3^k)$ &
\begin{minipage}{2.8cm}
${\rm Vect}(1,z_3,\dots,z_3^{k-2})$\\
$\dim=k-1$
\end{minipage}\\
\hline
 $D_k$ & \begin{minipage}{4cm} \tsep{1ex}$e_1=x^{2k+1}y+(-1)^{k+1}xy^{2k+1}$\\
$e_2=x^{2k}+(-1)^ky^{2k}$
\\ $e_3=x^2y^2$ \bsep{1ex}
\end{minipage} &
\begin{minipage}{4.6cm}
$\lambda_k((-1)^kz_1^2+(-1)^{k+1}z_2^2z_3+4z_3^{k+1})$
\\ with $\lambda_k=2k(-1)^{k+1}$
\end{minipage} &
\begin{minipage}{3.2cm}
${\rm Vect}(1,z_2,z_3,\dots,z_3^{k-1})$
\\ $\dim=k+1$
\end{minipage}\\
\hline
  $E_6$ &
\begin{minipage}{4.5cm} \tsep{1ex}$e_1=33y^8x^4-y^{12}+33y^4x^8-x^{12}$ \\
$e_2=14y^4x^4+x^8+y^8$\\
$e_3=x^5y-xy^5$ \bsep{1ex}
\end{minipage} & $4(z_1^2-z_2^3+108z_3^4)$ &
\begin{minipage}{4cm}
${\rm Vect}(1,z_2,z_3,z_2z_3,z_2z_3^2,z_3^2)$ \\
$\dim=6$
\end{minipage}\\
\hline
  $E_7$ &
\begin{minipage}{5cm} \tsep{1ex} $e_1=-34x^5y^{13}-yx^{17}+34y^5x^{13}+xy^{17}$\\
$e_2=-3y^{10}x^2+6y^6x^6-3y^2x^{10}$\\
$e_3=14y^4x^4+x^8+y^8$ \bsep{1ex}
\end{minipage} & $8(3z_1^2-12z_2^3+z_2z_3^3)$ &
\begin{minipage}{4.2cm}
${\rm Vect}(1,z_2,z_2^2,z_3,z_2z_3,z_2^2z_3,z_3^2)$ \\
$\dim=7$
\end{minipage}\\
\hline
  $E_8$ & \begin{minipage}{4.6cm}\tsep{1ex}
 $e_1=x^{30}+522x^{25}y^5-10\,005x^{20}y^{10}$\\
  $\phantom{e_1=}{} -10\,005x^{10}y^{20}-522x^5y^{25}+y^{30}  $ \\
 $e_2=x^{20}-228x^{15}y^5+494x^{10}y^{10}$\\
  $\phantom{e_2=}{} +228x^5y^{15}+y^{20}$ \\
 $e_3=x^{11}y+11x^6y^6-xy^{11}$ \bsep{1ex}
\end{minipage}
 & $10(-z_1^2+z_2^3+1\,728z_3^5)$ &
\begin{minipage}{2.8cm}
${\rm Vect}(z_2^iz_3^j)_{\substack{i=0,1,\\ j=0,\dots, 3}}$ \\
$\dim=8$
\end{minipage} \\
\hline
\end{tabular}}
\end{scriptsize}

\vspace{2mm}

Before carrying on with our study, we make a
digression in order to draw a parallel bet\-ween the Poisson and the
Hochschild
cohomologies of Klein surfaces, by recalling the result of A.~Pichereau.

\begin{theorem}[Pichereau]\label{Pich}
Consider the Poisson bracket defined on
$\mathbb{C}[z_1,z_1,z_3]$ by
\[\{\cdot,\cdot\}_{f}=\partial_3 f
\partial_1\wedge\partial_2+\partial_1 f
\partial_2\wedge\partial_3+\partial_2 f
\partial_3\wedge\partial_1=i(df)(\partial_1\wedge\partial_2\wedge\partial_3),
\] where $i$ is the contraction of
a multiderivation by a differential form. Denote by $HP^*_{f}$
(resp.\ $HP_*^{f}$) the Poisson cohomology (resp.\ homology) for
this bracket. Under the previous assumptions, the Poisson
cohomology $HP^*_{f}$ and the Poisson homology $HP_*^{f}$ of
$(\mathbb{C}[z_1,z_1,z_3]\,/\,\langle f\rangle,\{\cdot,\cdot\}_{f})$ are given by
\begin{gather*}
 HP^0_{f}=\mathbb{C},\qquad  HP^1_{f}\simeq HP^2_{f}=\{0\}, \\
 HP_0^{f}\simeq HP_2^{f}\simeq
\mathbb{C}[z_1,z_2,z_3]\,/\,\langle\partial_1 f,
\partial_2 f, \partial_3 f\rangle, \\
  \dim(HP_1^{f})=\dim(HP_0^{f})-1, \\
HP_j^{f}=HP_{f}^j=\{0\}\quad \text{if}\  \ j\geq 3.
\end{gather*}
\end{theorem}

The algebra $\mathbb{C}[x,y]$ is a Poisson algebra for
the standard symplectic bracket $\{\cdot,\cdot\}_{\rm std}$. As $G$ is
a subgroup of the symplectic group $\mathbf{Sp}_2\mathbb{C}$
(since $\mathbf{Sp}_2\mathbb{C}=\mathbf{SL}_2\mathbb{C}$), the
invariant algebra $\mathbb{C}[x,y]^G$ is a Poisson subalgebra of
$\mathbb{C}[x,y]$. The following proposition allows us to
deduce, from Theorem~\ref{Pich}, the Poisson cohomology and
homology of $\mathbb{C}[x,y]^G$ for the standard symplectic
bracket.

\begin{proposition}
With the choice made in the preceding table for the polynomial
$f$, the isomorphism of associative algebras
\begin{gather*}
  \pi: \ (\mathbb{C}[x,\,y]^G,
\{\cdot,\cdot\}_{\rm std})   \rightarrow
(\mathbb{C}[z_1, z_1, z_3]/\langle f\rangle,
\{\cdot,\cdot\}_{f}), \qquad  e_j   \mapsto  \overline{z_j}
\end{gather*}
is a Poisson isomorphism.
\end{proposition}

  In the sequel, we will calculate the Hochschild
cohomology of $\mathbb{C}[z_1,z_1,z_3]/\langle f\rangle$, and
we will immediately deduce the Hochschild cohomology of
$\mathbb{C}[x,y]^G$, with the help of the isomorphism~$\pi$.
Note that the fact that $\pi$ preserves the Poisson structures has
no incidence on the computation of the Hochschild cohomology.
Therefore, so as to simplify the calculations, we may replace the
polynomial $f$ by a simpler one, given in the following table
$$
\begin{array}{|c||c|c|c|c|c|}\hline
 \tsep{0.5mm}\bsep{0.5mm} G & A_k & D_k & E_6 & E_7 & E_8  \\ \hline
 \tsep{0.5mm}\bsep{0.5mm}  f & z_1^2+z_2^2+z_3^{k} & z_1^2+z_2^2z_3+z_3^{k} & z_1^2+z_2^3+z_3^4 & z_1^2+z_2^3+z_2z_3^3 & z_1^2+z_2^3+z_3^5
  \\\hline
\end{array}$$

 Indeed, the linear maps def\/ined by
 \begin{gather*}
  \mathbb{C}[\mathbf{z}]  \rightarrow   \mathbb{C}[\mathbf{z}], \\
  (z_1, z_2, z_3)   \mapsto   (\alpha_1z_1, \alpha_2z_2, \alpha_3z_3), \\
  (z_1, z_2, z_3)   \mapsto   (\alpha_1(z_1+z_2), \alpha_2(z_1+z_2), \alpha_3z_3)
\end{gather*}
are isomorphisms of associative algebras.

\subsection{Description of the cohomology spaces}\label{sec4.2}

We consider now the case
$A:=\mathbb{C}[z_1,z_2,z_3],/\,\langle f\rangle$ and we want
to calculate the Hochschild cohomology of $A$. We use the
notations of Section~\ref{sec2.1}, but we change the ordering of the basis:
we shall take $(\eta_1\eta_2, \eta_2\eta_3, \eta_3\eta_1)$
instead of $(\eta_1\eta_2, \eta_1\eta_3, \eta_2\eta_3)$. The
dif\/ferent spaces of the complex are now given by
\begin{gather*}
  \widetilde{T}(0)=A,\\
  \widetilde{T}(1)=A\eta_1\oplus A\eta_2\oplus A\eta_3, \\
  \widetilde{T}(2)=A b_1\oplus A\eta_1\eta_2\oplus A\eta_2\eta_3\oplus A\eta_3\eta_1, \\
  \widetilde{T}(3)=A b_1\eta_1\oplus A b_1\eta_2\oplus A b_1\eta_3\oplus A \eta_1\eta_2\eta_3, \\
  \widetilde{T}(4)=A b_1^2\oplus A b_1\eta_1\eta_2\oplus A b_1\eta_2\eta_3\oplus A b_1\eta_3\eta_1,  \\
\widetilde{T}(5)=A b_1^2\eta_1\oplus A b_1^2\eta_2\oplus A b_1^2\eta_3\oplus A b_1\eta_1\eta_2\eta_3,
\end{gather*}
i.e., for an arbitrary $p\in \mathbb{N}^*$,
\[
\widetilde{T}(2p)=A
b_1^p\oplus A b_1^{p-1}\eta_1\eta_2\oplus A
b_1^{p-1}\eta_2\eta_3\oplus A b_1^{p-1}\eta_3\eta_1,
\]
and
\[
\widetilde{T}(2p+1)=A b_1^p
\eta_1\oplus A b_1^p\eta_2\oplus A b_1^p\eta_3\oplus A b_1^{p-1}\eta_1\eta_2\eta_3.
\]

We have
\[
\frac{\partial}{\partial\eta_1}(\eta_1\wedge\eta_2\wedge\eta_3)=1\wedge\eta_2\wedge\eta_3=\eta_2\wedge\eta_3\wedge
1,
\]
 thus
 \[
 d_{\widetilde{T}}^{(3)}(\eta_1\eta_2\eta_3)=
\frac{\partial f}{\partial z_1} b_1\eta_2\eta_3+\frac{\partial
f}{\partial z_2} b_1\eta_3\eta_1
+\frac{\partial f}{\partial z_3} b_1\eta_1\eta_2.
\]

The matrices of $d_{\widetilde{T}}$ are therefore given
by
\begin{gather*}
{\rm Mat}_{\mathcal{B}_{1},\mathcal{B}_{2}}(d_{\widetilde{T}}^{(1)})=\left(
\begin{array}{ccc}
  \partial_{1}f & \partial_{2}f & \partial_{3}f \\
  0 & 0 & 0 \\
    0 & 0 & 0 \\
      0 & 0 & 0
\end{array}
\right), \\
\forall\; p\in \mathbb{N}^*, \ \ \
{\rm Mat}_{\mathcal{B}_{2p},\mathcal{B}_{2p+1}}(d_{\widetilde{T}}^{(2p)})=\left(
\begin{array}{cccc}
0 & \partial_{2}f & 0 & -\partial_{3}f \\
  0 & -\partial_{1}f & \partial_{3}f & 0 \\
    0 & 0 & -\partial_{2}f & \partial_{1}f \\
      0 & 0 & 0  & 0
\end{array}
\right),
\\ \forall\; p\in \mathbb{N}^*, \ \ \
{\rm Mat}_{\mathcal{B}_{2p+1},\mathcal{B}_{2p+2}}(d_{\widetilde{T}}^{(2p+1)})=\left(
\begin{array}{cccc}
\partial_{1}f & \partial_{2}f & \partial_{3}f & 0 \\
  0 & 0 & 0 & \partial_{3}f\\
    0 & 0 & 0  & \partial_{1}f\\
      0 & 0 & 0  & \partial_{2}f
\end{array}
\right).
\end{gather*}

We deduce
\begin{gather*}
  H^0   =  A, \\
  H^1  =  \{g_1\eta_1+g_2\eta_2+g_3\eta_3\, /\, (g_1, g_2,g_3)\in A^3\
\textmd{and}\
g_1\partial_{1}f+g_2\partial_{2}f+g_3\partial_{3}f=0\}\\
\phantom{H^1}{} \simeq   \left\{\mathbf{g}=\left(
\begin{array}{c}
  g_1 \\
  g_2 \\
  g_3
\end{array}
\right)\in A^3\, /\, \mathbf{g}\cdot\nabla f=0\right\}, \\
 H^{2}   =   \frac{\left\{g_0
b_1+g_3\eta_1\eta_2+g_1\eta_2\eta_3+g_2\eta_3\eta_1\, \Big/\,
\begin{array}{l} (g_0, g_1, g_2,g_3)\in A^4\ \textmd{and}\\
g_3 \partial_{2}f-g_2 \partial_{3}f=g_1 \partial_{3}f-g_3 \partial_{1}f\\
=g_2 \partial_{1}f-g_1 \partial_{2}f=0\end{array}\right\}}
{\{(g_1 \partial_{1}f+g_2 \partial_{2}f+g_3 \partial_{3}f)b_1 \,,
/\, (g_1, g_2, g_3)\in A^3\}} \\
\phantom{H^{2}}{}  \simeq   \left\{\mathbf{g}=\left(
\begin{array}{c}
 g_0\\
  g_1 \\
  g_2 \\
 g_3
\end{array}
\right)\in A^4\, \Big/\, \nabla f\wedge \left(
\begin{array}{c}
  g_1 \\
  g_2 \\
 g_3
\end{array}
\right)=0\right\}   \Bigg/   \left\{\left(
\begin{array}{c}
  \mathbf{g}\cdot\nabla f \\
  \mathbf{0}_{3,1}
\end{array}
\right)\, /\, \mathbf{g}\in A^3\right\} \\
\phantom{H^{2}}{} \simeq  \frac{A}{\langle \partial_{1}f, \partial_{2}f, \partial_{3}f\rangle_A}\oplus \{\mathbf{g}\in A^3\, /\, \nabla f\wedge \mathbf{g}=0\} .
\end{gather*}

For $p\geq 2,$
\begin{gather*}
  H^{2p}   =   \frac{\left\{g_0 b_1^p+g_3b_1^{p-1}\eta_1\eta_2+g_1b_1^{p-1}\eta_2\eta_3+g_2b_1^{p-1}\eta_3\eta_1
\Big/ \begin{array}{l} (g_0, g_1, g_2,g_3)\in A^4\ \textmd{and} \\
g_3\partial_{2}f-g_2 \partial_{3}f\\
=g_1 \partial_{3}f-g_3 \partial_{1}f\\
=g_2 \partial_{1}f-g_1 \partial_{2}f=0 \end{array}\right\}}
{\left\{\begin{array}{l}(g_1 \partial_{1}f+g_2 \partial_{2}f+g_3 \partial_{3}f)b_1^p\\
{}+g_0(\partial_{3}f b_1^{p-1}\eta_1\eta_2
+\partial_{1}f b_1^{p-1}\eta_2\eta_3+\partial_{2}f b_1^{p-1}\eta_3\eta_1)\end{array}\Big/ (g_0, g_1, g_2, g_3)\in A^3\right\}} \\
\phantom{H^{2p}}{} \simeq   \left\{\mathbf{g}=\left(
\begin{array}{c}
 g_0\\
  g_1 \\
  g_2 \\
 g_3
\end{array}
\right)\in A^4 \Big/ \nabla f\wedge \left(
\begin{array}{c}
  g_1 \\
  g_2 \\
 g_3
\end{array}
\right)=0\right\}   \! \Bigg/  \!  \left\{\left(
\begin{array}{c}
  \mathbf{g}\cdot\nabla f \\
  g_0\,\partial_{1}f \\
  g_0\,\partial_{2}f \\
  g_0\,\partial_{3}f
\end{array}
\right)  /\, \mathbf{g}\in A^3\ \textmd{and}\ g_0\in A\right\}
\\
\phantom{H^{2p}}{} \simeq  \frac{A}{\langle \partial_{1}f,  \partial_{2}f,  \partial_{3}f\rangle_A}\oplus
     \frac{\{\mathbf{g}\in A^3\, /\, \nabla f\wedge \mathbf{g}=0\}}{\left\{g\nabla f\, /\, g\in A\right\}}.
\end{gather*}

For $p\in \mathbb{N}^*$,
\begin{gather*}
  H^{2p+1}  =  \frac{\left\{g_1 b_1^p\eta_1+g_2 b_1^p\eta_2+g_3 b_1^p\eta_3+g_0b_1^{p-1}\eta_1\eta_2\eta_3\
\Big/ \begin{array}{l} (g_0,g_1,g_2g_3)\in A^4\ \textmd{and}\\
g_1\partial_{1}f+g_2\partial_{2}f+g_3\partial_{3}f=0,
\\ g_0 \partial_{3}f=g_0 \partial_{1}f=g_0 \partial_{2}f=0\end{array}\right\}}
{\left\{\begin{array}{l} (g_3 \partial_{2}f-g_2 \partial_{3}f)b_1^p\eta_1+(g_1 \partial_{3}f-g_3 \partial_{1}f)b_1^p\eta_2 \\
+(g_2 \partial_{1}f-g_1 \partial_{2}f)b_1^p\eta_3\end{array}\Big/
(g_1,g_2,g_3)\in A^3\right\}} \\
\phantom{H^{2p+1}}{}  \simeq  \left\{\left(
\begin{array}{c}
 g_1\\
  g_2 \\
  g_3 \\
 g_0
\end{array}
\right)\in A^4\! \Bigg/\! \begin{array}{l}
  \nabla f\cdot \left(
\begin{array}{c}
  g_1 \\
  g_2 \\
 g_3
\end{array}
\right)=0 \\
  g_0\,\partial_{3}f=g_0\,\partial_{1}f=g_0\,\partial_{2}f=0 \\
\end{array}\!
        \right\} \Bigg/   \left\{\left(
\begin{array}{c}
  \nabla f \wedge \mathbf{g} \\
  0 \\
\end{array}
\right) \Big/ \mathbf{g}\in A^3\right\}  \\
\phantom{H^{2p+1}}{} \simeq  \frac{\left\{\mathbf{g}\in A^3\, / \, \nabla f\cdot \mathbf{g}=0\right\}}
    {\left\{\nabla f\wedge \mathbf{g}\, /\, \mathbf{g}\in A^3\right\}}\oplus\{g\in A\, /\,
    g\partial_{3}f=g\partial_{1}f=g\partial_{2}f=0\}.
\end{gather*}

The following section will allow us to make those various spaces more explicit.

\subsection[Explicit calculations in the particular case where $f$ has separate variables]{Explicit calculations in the particular case where $\boldsymbol{f}$ has separate variables}\label{sec4.3}

In this section, we consider the polynomial
$f=a_1z_1^i+a_2z_2^j+a_3z_3^k$, with $2\leq i\leq j\leq k$ and $a_j\in \mathbb{C}^*$.
Its partial derivatives are $\partial_1f=ia_1z_1^{i-1}$,
$\partial_2f=ja_2z_2^{j-1}$ and
$\partial_3f=ka_3z_3^{k-1}$.

We already have
\[
H^0=\mathbb{C}[z_1,z_2,z_3]\,/\,\langle
a_1z_1^i+a_2z_2^j+a_3z_3^k\rangle.\]

Moreover, as $f$ is weighted homogeneous,
Euler's formula gives
\[
\frac{1}{i}z_1\partial_1f+\frac{1}{j}z_2\partial_2f+\frac{1}{k}z_3\partial_3f=f.
\]
So we have the inclusion $\langle f\rangle\subset\langle
\partial_1f, \partial_2f, \partial_3f\rangle$, thus
\begin{gather*} \frac{A}
{\langle
\partial_1f,\partial_2f,\partial_3f\rangle_A}\simeq
\frac{\mathbb{C}[z_1,z_2,z_3]}{\langle
\partial_1f,\partial_2f,\partial_3f\rangle}\\
\phantom{\frac{A}
{\langle
\partial_1f,\partial_2f,\partial_3f\rangle_A}}{} \simeq
{\rm Vect}\left(z_1^pz_2^qz_3^r\, /\, p\in\llbr 0,i-2\rrbr,\ q\in\llbr
0,j-2\rrbr,\
r\in\llbr 0,k-2\rrbr\right).
\end{gather*}
Finally, as $\partial_1f$ and $f$ are relatively prime, if $g\in
A$ verif\/ies $g\partial_1f=0 \mod \langle f\rangle$, then $g\in
\langle f\rangle$, i.e.~$g$ is zero in $A$.

Now we determine the set
\[
\left\{\mathbf{g}=\left(
\begin{array}{c}
  g_1 \\
  g_2 \\
  g_3
\end{array}
\right)\in A^3\, /\, \mathbf{g}\cdot\nabla f=0\right\}.
\]
First we have
\[
\langle
f, \partial_1f, \partial_2f\rangle=\langle
a_1z_1^i+a_2z_2^j+a_3z_3^k, z_1^{i-1}, z_2^{j-1}\rangle=\langle
z_1^{i-1}, z_2^{j-1}, z_3^k\rangle.
\] Thus the only monomials
which are not in this ideal are the elements $z_1^pz_2^qz_3^r$
with $p\in\llbr
0, i-2\rrbr$, $q\in\llbr 0, j-2\rrbr$, and $r\in\llbr 0, k-1\rrbr$.

So every polynomial $P\in \mathbb{C}[\mathbf{z}]$ may be written
in the form  \[ P=\alpha f+\beta
\partial_1f+\gamma
\partial_2f+\sum_{\substack{p=0,\dots, i-2 \\ q=0,\dots,
j-2 \\ r=0,\dots,
k-1}}a_{pqr}z_1^pz_2^qz_3^r .\]
The polynomials $P\in \mathbb{C}[\mathbf{z}]$ such that
$P\partial_3f\in \langle f, \partial_1f, \partial_2f\rangle$ are
therefore the following ones
\[ P=\alpha f+\beta
\partial_1f+\gamma
\partial_2f+\sum_{\substack{p=0,\dots, i-2 \\ q=0,\dots,
j-2 \\ r=1,\dots, k-1}}a_{pqr}z_1^pz_2^qz_3^r .
\]
So we have calculated ${\rm Ann}_{\langle f, \partial_1f, \partial_2f\rangle}(\partial_3f)$.
The equation
\begin{gather*}
\mathbf{g}\cdot\nabla f=0 \mod \langle f\rangle
\end{gather*}
leads to $g_3\in {\rm Ann}_{\langle
f, \partial_1f, \partial_2f\rangle}(\partial_3f)$, i.e.
\begin{gather*}
g_3=\alpha f+\beta\partial_1f+\gamma
\partial_2f+\sum_{\substack{p=0,\dots, i-2 \\ q=0,\dots,
j-2 \\ r=1\dots k-1}}a_{pqr}z_1^pz_2^qz_3^r,
\end{gather*}
with $(\alpha,\beta,\gamma)\in \mathbb{C}[\mathbf{z}]^3$.
Hence
\begin{gather*}
g_2\partial_2f+\gamma\partial_2f\partial_3f+\sum_{\substack{p=0,\dots,
i-2 \\ q=0,\dots, j-2 \\ r=1,\dots, k-1}}a_{pqr}z_1^pz_2^qz_3^r
\partial_3f\in \langle f,\partial_1f\rangle.
\end{gather*}
Thus, according to Euler's formula,
\begin{gather*}
\partial_2f\left(g_2+\gamma\partial_3f-\frac{k}{j}\sum_{\substack{p=0,\dots,
i-2 \\ q=0\dots j-2 \\ r=1,\dots,
k-1}}a_{pqr}z_1^pz_2^{q+1}z_3^{r-1}\right)\in \langle
f,\,\partial_1f\rangle.
\end{gather*}
Since ${\rm Ann}_{\langle f,\partial_1f\rangle}(\partial_2f)=\langle
f,\,\partial_1f\rangle$, this equation is equivalent to
\[g_2=-\gamma\partial_3f+\frac{k}{j}\sum_{\substack{p=0,\dots, i-2 \\
q=0,\dots, j-2 \\ r=1,\dots,
k-1}}a_{pqr}z_1^pz_2^{q+1}z_3^{r-1}+\delta
f+\varepsilon\partial_1f,
\] with $\delta,\varepsilon\in
\mathbb{C}[\mathbf{z}]$. It follows that
\begin{gather*}
g_1\partial_1f+\beta\partial_1f\partial_3f+\varepsilon\partial_1f\partial_2f\nonumber\\
\qquad{}+\sum_{\substack{p=0,\dots,
i-2 \\ q=0,\dots, j-2 \\ r=1,\dots, k-1}}a_{pqr}z_1^pz_2^qz_3^r
\partial_3f+\frac{k}{j}\sum_{\substack{p=0,\dots,
i-2 \\ q=0,\dots, j-2 \\ r=1,\dots,
k-1}}a_{pqr}z_1^pz_2^{q+1}z_3^{r-1}
\partial_2f\in \langle f\rangle.
\end{gather*}
And, according to Euler's formula,
\begin{gather*}
\partial_1f\left(g_1+\beta\partial_3f+\varepsilon\partial_2f-\frac{k}{i}\sum_{\substack{p=0,\dots,
i-2 \\ q=0,\dots, j-2 \\ r=1,\dots,
k-1}}a_{pqr}z_1^{p+1}z_2^qz_3^{r-1}\right)\in \langle f\rangle,
\end{gather*}
i.e.
\begin{gather*}
g_1=-\beta\partial_3f-\varepsilon\partial_2f+\frac{k}{i}\sum_{\substack{p=0,\dots,
i-2 \\ q=0,\dots, j-2 \\ r=1,\dots,
k-1}}a_{pqr}z_1^{p+1}z_2^qz_3^{r-1}+\eta f,
\end{gather*}
with $\eta\in \mathbb{C}[\mathbf{z}]$. Finally
\begin{gather*}
\left\{\mathbf{g}\in A^3\, /\,
\mathbf{g}\cdot\nabla f=0 \right\}\\
\quad{}=\left\{ \nabla f\wedge\left(%
\begin{array}{c}
  -\gamma \\
  \beta \\
   -\varepsilon
\end{array}
\right)+\sum_{\substack{p=0,\dots, i-2 \\ q=0,\dots, j-2 \\ r=1,\dots,
k-1}}a_{pqr}z_1^pz_2^qz_3^{r-1}\left(%
\begin{array}{c}
  \frac{k}{i}z_1 \\
  \frac{k}{j}z_2 \\
  z_3
\end{array}%
\right) \Big/
  (\beta,\gamma,\varepsilon)\in A^3 \
  \textrm{and}\ a_{pqr}\in
\mathbb{C}
 \right\}.
\end{gather*}
We deduce immediately the cohomology spaces of odd degrees
\begin{alignat*}{3}
& \forall\; p\geq 1,\quad &&  H^{2p+1}  \simeq  \mathbb{C}^{(i-1)(j-1)(k-1)},& \\
 &&& H^{1}  \simeq   \nabla f\wedge\left(\mathbb{C}[\mathbf{z}]\,/\,\langle f \rangle\right)^3\oplus\mathbb{C}^{(i-1)(j-1)(k-1)}.
\end{alignat*}

It remains to determine the set
\[
\left\{\mathbf{g}=\left(
\begin{array}{c}
  g_1 \\
  g_2 \\
  g_3
\end{array}
\right)\in A^3\, /\, \nabla f\wedge\mathbf{g}=0\right\}.
\]
Let $\mathbf{g}\in A^3$ be such that $\nabla f\wedge\mathbf{g}=0$.
This means that, modulo $\langle f\rangle$, $\mathbf{g}$ verif\/ies
the system \begin{gather*}
  \partial_2f g_3-\partial_3f g_2  =  0, \qquad
  \partial_3f g_1-\partial_1f g_3  =  0, \qquad
  \partial_1f g_2-\partial_2f g_1  =  0.
\end{gather*}
The f\/irst equation gives, modulo $\langle
f,\partial_2f\rangle$, $\partial_3fg_2=0$.
Now ${\rm Ann}_{\langle f,\partial_2f\rangle}(\partial_3f)=\langle
f,\partial_2f\rangle$, therefore $g_2=\alpha f+\beta \partial_2f$. Hence
\[
\partial_2f(g_3-\beta\partial_3f)=0\mod \langle f\rangle,
\]
i.e.\ $g_3=\gamma f+\beta\partial_3f$.
Finally, we obtain{\samepage
\[
\partial_3f(g_1-\beta\partial_1f)=0\mod \langle f\rangle,
\]
i.e.\ $g_1=\delta f+\beta\partial_1f$.
So, $\{\mathbf{g}\in A^3\, /\, \nabla
f\wedge\mathbf{g}=0\}=\left\{\beta\nabla f\, /\, \beta\in
A\right\}$.}

We deduce the cohomology spaces of even degrees (for the
direct sum, we use the same argument as in Section \ref{sec3.2})
\begin{alignat*}{3}
& \forall\; p\geq 2,\quad &&  H^{2p}   \simeq   A\,
/\, \langle
\partial_1f, \partial_2f, \partial_3f\rangle\simeq
\mathbb{C}[\mathbf{z}]\, /\, \langle
z_1^{i-1},z_2^{j-1},z_3^{k-1}\rangle & \\
 &&& \phantom{H^{2p}}{} \simeq  {\rm Vec}t\left(z_1^pz_2^qz_3^r\, /\, p\in\llbr 0,i-2\rrbr,\ q\in\llbr0,j-2\rrbr,
 r\in\llbr0,k-2\rrbr\right)  & \\
&&& \phantom{H^{2p}}{}  \simeq  \mathbb{C}^{(i-1)(j-1)(k-1)},& \\
& && H^{2}  \simeq   \{\beta\,\nabla f\, /\, \beta\in
A\}\oplus\mathbb{C}^{(i-1)(j-1)(k-1)} & \\
 &&& \phantom{H^{2}}{}  \simeq  \mathbb{C}[\mathbf{z}]\, /\, \langle
a_1z_1^i+a_2z_2^j+a_3z_3^k\rangle\oplus\mathbb{C}^{(i-1)(j-1)(k-1)}.&
\end{alignat*}

\begin{remark} We also have \[
\nabla
f\wedge\left(\mathbb{C}[\mathbf{z}]\,/\,\langle f
\rangle\right)^3\simeq \left(\mathbb{C}[\mathbf{z}]\,/\,\langle f
\rangle\right)^3  /\, \{\mathbf{g}\, /\, \nabla f\wedge
\mathbf{g}=0\}=\left(\mathbb{C}[\mathbf{z}]\,/\, \langle f
\rangle\right)^3  /\, (\mathbb{C}[\mathbf{z}]\,/\, \langle f
\rangle)\nabla f.\]
Moreover the map
\begin{gather*}
  (\mathbb{C}[\mathbf{z}]\,/\,\langle f\rangle)^2   \rightarrow   \nabla f\wedge\left(\mathbb{C}[\mathbf{z}]\,/\,\langle f \rangle\right)^3, \qquad
  \left(\begin{array}{c}
  g_1 \\
  g_2
\end{array}
\right)  \mapsto  \nabla f\wedge\left(%
\begin{array}{c}
  g_1 \\
  g_2 \\
  0
\end{array}%
\right)
\end{gather*} is injective, thus  $\nabla f\wedge\left(\mathbb{C}[\mathbf{z}]\,/\,\langle f
\rangle\right)^3$ is inf\/inite-dimensional.
\end{remark}

\begin{remark}
In particular, we obtain the cohomology for the cases where
$f=z_1^2+z_2^2+z_3^{k}$, $f=z_1^2+z_2^3+z_3^4$ and
$f=z_1^2+z_2^3+z_3^5$. These cases correspond respectively to
the types $A_k$, $E_6$ and $E_8$ of the Klein surfaces.
\end{remark}

The following table sums up the results of those three
special cases:
$$
\begin{array}{|@{\,}l@{\,}||@{\,\,}l@{\,}|@{\,}l@{\,}|@{\,\,}l@{\,}|@{\,}l@{\,}|}
  \hline
\tsep{0.5mm}\bsep{0.5mm}    & H^0 & H^1 & H^2 & H^{p},\ p\geq 3 \\
  \hline\hline
\tsep{0.5mm}\bsep{0.5mm}    A_k & \mathbb{C}[\mathbf{z}]\, /\, \langle z_1^2+z_2^2+z_3^{k}\rangle & \nabla f\wedge\left(\mathbb{C}[\mathbf{z}]\, /\, \langle f\rangle\right)^3\oplus \mathbb{C}^{k-1} & \mathbb{C}[\mathbf{z}]\, /\, \langle z_1^2+z_2^2+z_3^{k}\rangle\oplus \mathbb{C}^{k-1} &  \mathbb{C}^{k-1}
  \\ \hline
\tsep{0.5mm}\bsep{0.5mm}    E_6 & \mathbb{C}[\mathbf{z}]\, /\, \langle z_1^2+z_2^3+z_3^4\rangle & \nabla f\wedge\left(\mathbb{C}[\mathbf{z}]\, /\, \langle f\rangle\right)^3\oplus \mathbb{C}^6 & \mathbb{C}[\mathbf{z}]\, /\, \langle z_1^2+z_2^3+z_3^4\rangle\oplus \mathbb{C}^6 &  \mathbb{C}^6
  \\ \hline
\tsep{0.5mm}\bsep{0.5mm}    E_8 & \mathbb{C}[\mathbf{z}]\, /\, \langle z_1^2+z_2^3+z_3^5\rangle & \nabla f\wedge\left(\mathbb{C}[\mathbf{z}]\, /\, \langle f\rangle\right)^3\oplus \mathbb{C}^8 & \mathbb{C}[\mathbf{z}]\, /\, \langle z_1^2+z_2^3+z_3^5\rangle\oplus \mathbb{C}^8 &  \mathbb{C}^8  \\
  \hline
\end{array}
$$

The cases where $f=z_1^2+z_2^2z_3+z_3^{k}$ and
$f=z_1^2+z_2^3+z_2z_3^3$, i.e.\ respectively $D_k$ and $E_7$
are studied in the following section.

\subsection[Explicit calculations for $D_k$ and $E_7$]{Explicit calculations for $\boldsymbol{D_k}$ and $\boldsymbol{E_7}$}\label{sec4.4}

\subsubsection[Case of $f=z_1^2+z_2^2z_3+z_3^{k}$, i.e. $D_k$]{Case of $\boldsymbol{f=z_1^2+z_2^2z_3+z_3^{k}}$, i.e. $\boldsymbol{D_k}$}\label{sec4.4.1}

 In this section, we consider the polynomial
$f=z_1^2+z_2^2z_3+z_3^{k}$, with $k\geq 3$.
Its partial derivatives are $\partial_1f=2z_1$,
$\partial_2f=2z_2z_3$ and
$\partial_3f=z_2^2+kz_3^{k-1}$.

 We already have
\[
 H^0=\mathbb{C}[\mathbf{z}]\,/\,\langle
z_1^2+z_2^2z_3+z_3^{k}\rangle.
\]

 Besides, since $f$ is weighted homogeneous,
Euler's formula gives
\begin{gather}\label{euler3Dk}
 \frac{k}{2}z_1 \partial_1f+\frac{k-1}{2}z_2 \partial_2f+z_3 \partial_3f=kf  .
\end{gather}
Thus, we have the inclusion $\langle f\rangle\subset\langle
\partial_1f, \partial_2f, \partial_3f\rangle$.
Moreover, a Groebner basis of $\langle
\partial_1f, \partial_2f, \partial_3f\rangle$ is $[z_3^{k},z_2z_3,z_2^2+kz_3^{k-1},
z_1]$, therefore \[
\frac{A} {\langle
\partial_1f, \partial_2f, \partial_3f\rangle_A}\simeq
\frac{\mathbb{C}[z_1,z_2, z_3]}{\langle
\partial_1f, \partial_2f, \partial_3f\rangle}\simeq
{\rm Vect}\big(z_2, 1, z_3,\dots, z_3^{k-1}\big).\]
Finally, as $\partial_1f$ and $f$ are relatively prime, if $g\in
A$ verif\/ies $g\partial_1f=0 \mod \langle f\rangle$, then $g\in
\langle f\rangle$, i.e.~$g$ is zero in $A$, thus $\{g\in A\, /\, g \partial_{3}f=g \partial_{1}f=g\,\partial_{2}f=0\}=0$.

 Now we determine the set
\[
\left\{\mathbf{g}=\left(
\begin{array}{c}
  g_1 \\
  g_2 \\
  g_3
\end{array}
\right)\in A^3\, /\, \mathbf{g}\cdot\nabla f=0\right\}.
\]
A Groebner basis of $\langle f, \partial_1f, \partial_3f\rangle$
is $[z_1,  z_3^{k},  z_2^2+kz_3^{k-1}]$, thus a basis of
$\mathbb{C}[\mathbf{z}]\, /\, \langle
f, \partial_1f, \partial_3f\rangle$ is $\{z_2^iz_3^j\, /\,
i\in\{0, 1\},\ j\in\llbr
0, k-1\rrbr\}$.
We have already solved the equation $p \partial_2f=0$ in this
space; the solutions of this equation in $\mathbb{C}[\mathbf{z}]\,
/\, \langle f, \partial_1f, \partial_3f\rangle$ are of the form
\[
p=a_{0,k-1}z_3^{k-1}+\sum_{j=0}^{k-1}a_{1,j}z_2z_3^j,
\]
where $a_{0,k-1}, a_{1,j}\in \mathbb{C}$.

Let $\mathbf{g}=\left(
\begin{array}{c}
  g_1 \\
  g_2 \\
  g_3
\end{array}%
\right)\in A^3$ satisfy the equation
\begin{gather*}
\mathbf{g}\cdot\nabla f=0 \mod \langle f\rangle.
\end{gather*}
Then we have
\begin{gather*}
g_2\partial_2f=0 \mod \langle
f, \partial_1f, \partial_3f\rangle,
\end{gather*}
hence
\begin{gather*}
g_2=\alpha f+\beta\partial_1f+\gamma
\partial_3f+az_3^{k-1}+\sum_{j=0}^{k-1}b_jz_2z_3^j,
\end{gather*}
with $(\alpha, \beta, \gamma)\in \mathbb{C}[\mathbf{z}]^3$.
And
\begin{gather}\label{31Deq4}
g_3\partial_3f+\gamma\partial_3f\partial_2f+az_3^{k-1}\partial_2f+\sum_{j=0}^{k-1}b_jz_2z_3^j\partial_2f
\in \langle f, \partial_1f\rangle.
\end{gather}
 Now according to Euler's formula (\ref{euler3Dk}) and the
 equality
\begin{gather*}
 z_3^{k}z_2=\frac{1}{1-k}\left(z_2f-z_2z_3\partial_3f-\frac{1}{2}z_2z_1\partial_1f\right)=-\frac{1}{1-k}z_2z_3\partial_3f \mod \langle f, \partial_1f\rangle,
\end{gather*}
Equation (\ref{31Deq4}) becomes
\begin{gather*}
\partial_3f\left(g_3+\gamma\partial_2f-\frac{2a}{1-k}z_2z_3-\sum_{j=0}^{k-1}b_j\frac{2}{k-1}z_3^{j+1}\right)\in \langle
f, \partial_1f\rangle.
\end{gather*}
As ${\rm Ann}_{\langle f, \partial_1f\rangle}(\partial_3f)=\langle
f, \partial_1f\rangle$, this equation is equivalent to
\[
g_3=-\gamma\partial_2f+\frac{2a}{1-k}z_2z_3+\sum_{j=0}^{k-1}b_j\frac{2}{k-1}z_3^{j+1}+\delta
f+\varepsilon\partial_1f,
\] with
$\delta, \varepsilon\in \mathbb{C}[\mathbf{z}]$.
We f\/ind
\begin{gather*}
g_1=-\beta\partial_2f-\varepsilon\partial_3f+\sum_{j=0}^{k-1}b_j\frac{k}{k-1}z_1z_3^j+\frac{a}{1-k}z_2z_1+\eta
f,
\end{gather*}
with $\eta\in \mathbb{C}[\mathbf{z}]$.
 Finally, we have
\begin{gather*}
\left\{\mathbf{g}\in A^3\, /\,
\mathbf{g}\cdot\nabla f=0 \right\}\\
\quad{}=\left\{ \nabla f\wedge\left(%
\begin{array}{c}
  \gamma \\
  \varepsilon \\
   -\beta
\end{array}%
\right)+\sum_{j=0}^{k-1}b_j\left(%
\begin{array}{c}
  \frac{k}{k-1}z_1z_3^j \\
  z_2z_3^j \\
  -\frac{2}{1-k}z_3^{j+1}
\end{array}%
\right)+a\left(%
\begin{array}{c}
  \frac{1}{1-k}z_2z_1 \\
  z_3^{k-1} \\
  \frac{2}{1-k}z_2z_3
\end{array}%
\right)  \Bigg/  \begin{array}{l}
  (\beta, \gamma, \varepsilon)\in A^3 \\
  \textrm{and}\ a, b_j\in \mathbb{C} \\
\end{array} \right\},
\end{gather*}
as well as cohomology spaces of odd degrees (for
the direct sum, we use the same argument as in Section~\ref{sec3.2})
\begin{alignat*}{3}
& \forall\; p\geq 1,\quad && H^{2p+1}   \simeq   \mathbb{C}^{k+1}, & \\
&&& H^{1}   \simeq    \nabla f\wedge\left(\mathbb{C}[\mathbf{z}]\,/\,\langle f \rangle\right)^3\oplus\mathbb{C}^{k+1}.
\end{alignat*}

 To show $\{\mathbf{g}\in A^3\, /\, \nabla
f\wedge\mathbf{g}=0\}=\left\{f \mathbf{g}+\beta\nabla f\, /\,
\mathbf{g}\in
A^3,\ \beta\in A\right\}$, we proceed as in the case of separate variables.
We deduce the cohomology spaces of even degrees
\begin{alignat*}{3}
 & \forall\; p\geq 2,\quad &&  H^{2p}   \simeq   A\,
/\, \langle
\partial_1f, \partial_2f, \partial_3f\rangle\simeq
{\rm Vect}\left(z_2, 1, z_3,\dots, z_3^{k-1}\right) \simeq \mathbb{C}^{k+1}, & \\
&&& H^{2}   \simeq    \{\beta \nabla f\, /\, \beta\in
A\}\oplus\mathbb{C}^{k+1}\simeq \mathbb{C}[\mathbf{z}]\, /\, \langle
z_1^2+z_2^2z_3+z_3^{k}\rangle\oplus\mathbb{C}^{k+1}.
\end{alignat*}

\subsubsection[Case of $f=z_1^2+z_2^3+z_2z_3^3$, i.e. $E_7$]{Case of $\boldsymbol{f=z_1^2+z_2^3+z_2z_3^3}$, i.e.\ $\boldsymbol{E_7}$}\label{sec4.4.2}

 Here we have $\partial_1f=2z_1,
\partial_2f=3z_2^2+z_3^3$ and $\partial_3f=3z_2z_3^2$.
The proof is similar to that of the previous cases.
A~Groebner basis of $\langle
\partial_1f, \partial_2f, \partial_3f\rangle$ is $[z_3^5,  z_2z_3^2,  3z_2^2+z_3^3,
z_1]$.
Similarly, a~Groebner basis of $\langle
f, \partial_1f, \partial_2f\rangle$ is $[z_3^6, z_2z_3^3,  3z_2^2+z_3^3,  z_1]$.
We obtain the following results
\begin{alignat*}{3}
& \forall\; p\geq 1,\quad && H^{2p+1}   \simeq   \mathbb{C}^7,& \\
&&& H^{1}   \simeq    \nabla f\wedge\left(\mathbb{C}[\mathbf{z}]\,/\,\langle f \rangle\right)^3\oplus\mathbb{C}^7,&\\
&\forall\; p\geq 2,\quad && H^0   =   \mathbb{C}[\mathbf{z}]\, /\, \langle z_1^2+z_2^3+z_2z_3^3\rangle, & \\
&&& H^{2p}   \simeq   A\, /\, \langle
\partial_1f, \partial_2f, \partial_3f\rangle\simeq
{\rm Vect}\left(z_2, z_2^2, 1, z_3, z_3^2, z_3^3, z_3^4\right) \simeq \mathbb{C}^7,& \\
&&& H^{2}   \simeq    \{\beta \nabla f\, /\, \beta\in
A\}\oplus\mathbb{C}^7\simeq \mathbb{C}[\mathbf{z}]\, /\, \langle
z_1^2+z_2^3+z_2z_3^3\rangle\oplus\mathbb{C}^7. &
\end{alignat*}

\begin{remark}
In all the previously studied cases, there exists a triple
$(i, j, k)$ such that  $\{i, j, k\}=\{1, 2, 3\}$, and such
that the map
\begin{gather*}
  \mathbb{C}[\mathbf{z}]\, /\, \langle \partial_1 f, \partial_2 f, \partial_3 f
  \rangle   \rightarrow   \{g\in\mathbb{C}[\mathbf{z}]\,/\,\langle f,
  \partial_j f, \partial_k f \rangle\, /\, g \partial_i f=0 \}, \\
  P \mod \langle \partial_1 f, \partial_2 f,\partial_3 f
  \rangle   \mapsto   z_i P \mod \langle f, \partial_j f, \partial_k f \rangle
\end{gather*}
is an isomorphism of vector spaces.
\end{remark}

\subsection{Homology}\label{sec4.5}

 The study is the same as the one of the
Hochschild cohomology, and we proceed as in Section~\ref{sec3.5}.
Here, we have
\begin{alignat*}{3}
&&&  \widetilde{\Omega}(0)=A, \qquad   \widetilde{\Omega}(-1)=A\xi_1\oplus A\xi_2\oplus A\xi_3, &\\
& \forall\; p\in \mathbb{N}^*, \quad & & \widetilde{\Omega}(-2p)=Aa_1^p\oplus Aa_1^{p-1}\xi_1\xi_2\oplus Aa_1^{p-1}\xi_2\xi_3\oplus Aa_1^{p-1}\xi_3\xi_1, &\\
 & \forall\ p\in \mathbb{N}^*,  \quad && \widetilde{\Omega}(-2p-1)=Aa_1^p\xi_1\oplus Aa_1^{p}\xi_2\oplus Aa_1^{p}\xi_3\oplus Aa_1^{p-1}\xi_1\xi_2\xi_3.&
\end{alignat*}
This def\/ines the bases $\mathcal{V}_{p}$. The dif\/ferential is
$d_{\widetilde{\Omega}}=(\xi_1\partial_1f+\xi_2\partial_2f+\xi_3\partial_3f)\frac{\partial}{\partial
a_1}$.

By setting $Df:=\left(
  \partial_3f \ \partial_1f \ \partial_2f
\right)$, we deduce the matrices
\begin{alignat*}{3}
&&& {\rm   Mat}_{\mathcal{V}_{-2},\mathcal{V}_{-1}}\left(d_{\widetilde{\Omega}}^{(-2)}\right)=\left(%
\begin{array}{cc}
  \nabla f & \mathbf{0}_{3,3} \\
\end{array}%
\right),& \\
&  \forall\; p\geq 2,\quad && {\rm Mat}_{\mathcal{V}_{-2p},\mathcal{V}_{-2p+1}}\left(d_{\widetilde{\Omega}}^{(-2p)}\right)=\left(%
\begin{array}{cc}
  \nabla f & \mathbf{0}_{3,3} \\
  0 & (p-1)Df
\end{array}%
\right),& \\
&  \forall\; p\geq 1,\quad && {\rm Mat}_{\mathcal{V}_{-2p-1},\mathcal{V}_{-2p}}\left(d_{\widetilde{\Omega}}^{(-2p-1)}\right)=\left(%
\begin{array}{cccc}
  0 & 0 & 0 & 0 \\
  -p\partial_2f & p\partial_1f & 0 & 0 \\
  0 & -p\partial_3f & p\partial_2f & 0 \\
  p\partial_3f & 0 & -p\partial_1f & 0
\end{array}%
\right).
\end{alignat*}

The cohomology spaces read as
\begin{gather*}
  L^0   =   A,\qquad
  L^{-1}  =  \frac{A^3}{\{g\nabla f\, /\, g\in A\}},\\
  L^{-2}  =  \{g\in A\, /\, g\partial_1f=g\partial_2f=g\partial_3f=0\}
  \oplus \frac{A^3}{\{\nabla f\wedge \mathbf{g}\, /\, \mathbf{g}\in A^3\}}.
\end{gather*}

For $p\geq 2$,
\begin{gather*}
  L^{-2p}  \simeq  \{g\in A\, /\, g\partial_1f=g\partial_2f=g\partial_3f=0\}\oplus \frac{\{\mathbf{g}\in A^3\, /\, \mathbf{g}\cdot\nabla f=0\}}{\{\nabla f\wedge \mathbf{g}\, /\, \mathbf{g}\in A^3\}}.
\end{gather*}

For $p\in \mathbb{N}^*$,
\begin{gather*}
L^{-2p-1}   \simeq    \frac{\{\mathbf{g}\in A^3\, /\, \nabla f\wedge\mathbf{g}=0\}}{\{g\nabla f\, /\, g\in A\}}\oplus \frac{A}{\langle \nabla f\rangle_A}.
\end{gather*}

From now on, we assume that either $f$ has
separate variables, or $f$ is of type $D_k$ or $E_7$. Then we have
$\left\{g\in A\, /\,
g\partial_1f=g\partial_2f=g\partial_3f=0\right\}=\{0\}$, and
according to Euler's formula,
\[
\frac{A}{\langle\nabla
f\rangle_A}\simeq\frac{\mathbb{C}[\mathbf{z}]}{\langle\nabla
f\rangle}.
\]
Most of the spaces have already been computed in Sections~\ref{sec4.3} and~\ref{sec4.4}.
In particular, we have $A^3\,/\,A\nabla f\simeq \nabla f\wedge A^3$.
Moreover, $\{\nabla f\wedge \mathbf{g}\,/\,\mathbf{g}\in
A^3\}\subset\{\mathbf{g}\in A^3\,/\,\mathbf{g}\cdot\nabla f=0\}$,
thus \[
\dim\left(A^3\,/\,\{\nabla f\wedge
\mathbf{g}\,/\,\mathbf{g}\in A^3\}\right)\geq
\dim\left(A^3\,/\,\{\mathbf{g}\in A^3\,/\,\mathbf{g}\cdot\nabla
f=0\}\right).
\] And $A^3\,/\,\{\mathbf{g}\in
A^3\,/\,\mathbf{g}\cdot\nabla f=0\}\simeq \{\mathbf{g}\cdot\nabla
f\,/\,\mathbf{g}\in A^3\}$. Since the map{\samepage
\[
g\in A\mapsto \left(
\begin{array}{c}
  g \\
  0 \\
  0
\end{array}
\right)\cdot\nabla f\in \{\mathbf{g}\cdot\nabla
f\,/\,\mathbf{g}\in A^3\}
\] is injective, $A^3\,/\,\{\nabla
f\wedge \mathbf{g}\,/\,\mathbf{g}\in A^3\}$ is
inf\/inite-dimensional.}

In the following table we collect the results
for the Hochschild homology in the various cases
$$
\begin{array}{|l||l|l|l|l|}\hline
\tsep{0.5mm}\bsep{0.5mm} \textrm{Type} & HH_0=A & HH_1 & HH_2 & HH_p,\,p\geq 3\\
\hline\hline
\tsep{0.5mm}\bsep{0.5mm} A_k & \mathbb{C}[\mathbf{z}]\,/\,\langle
z_1^2+z_2^2+z_3^{k}\rangle & \nabla f\wedge A^3 & A^3\,/\,(\nabla f\wedge A^3)& \mathbb{C}^{k-1} \\
\hline
\tsep{0.5mm}\bsep{0.5mm} D_k & \mathbb{C}[\mathbf{z}]\,/\,\langle z_1^2+z_2^2z_3+z_3^{k}\rangle & \nabla f\wedge A^3 & A^3\,/\,(\nabla f\wedge A^3) & \mathbb{C}^{k+1} \\ \hline
\tsep{0.5mm}\bsep{0.5mm}  E_6 & \mathbb{C}[\mathbf{z}]\,/\,\langle z_1^2+z_2^3+z_3^4\rangle & \nabla f\wedge A^3 & A^3\,/\,(\nabla f\wedge A^3) & \mathbb{C}^{6} \\
\hline
\tsep{0.5mm}\bsep{0.5mm} E_7 & \mathbb{C}[\mathbf{z}]\,/\,\langle
z_1^2+z_2^3+z_2z_3^3\rangle & \nabla f\wedge A^3   &
A^3\,/\,(\nabla f\wedge A^3) & \mathbb{C}^{7} \\ \hline
\tsep{0.5mm}\bsep{0.5mm} E_8 & \mathbb{C}[\mathbf{z}]\,/\,\langle z_1^2+z_2^3+z_3^5\rangle & \nabla f\wedge A^3 & A^3\,/\,(\nabla f\wedge A^3) & \mathbb{C}^{8}  \\ \hline
\end{array}
$$

\subsection*{Acknowledgements}

\noindent I would like to thank my thesis advisors Gadi Perets and
Claude Roger for their ef\/f\/icient and likeable help, for their
great availability, and for the
time that they devoted to me all along this study.
I also thank Daniel Sternheimer who paid attention to my work and
the referees for their relevant remarks
and their judicious advice.
And I am grateful to Serge Parmentier for the rereading of my
English text.

\pdfbookmark[1]{References}{ref}
\LastPageEnding

\end{document}